\documentclass[apl,reprint,superscriptaddress]{revtex4-1}
\usepackage{bm,amsmath,amssymb,url,float,cases,fancybox,braket}
\usepackage[dvipdfmx]{xcolor}
\usepackage{graphicx}

\renewcommand{\H}{{\mathcal H}}

\newcommand{\bS}{{\bm{S}}}
\newcommand{\bM}{{\bm{M}}}

\newcommand{\bW}{{\bm{W}}}
\newcommand{\bk}{{\bm{k}}}
\newcommand{\br}{{\bm{r}}}
\newcommand{\bE}{{\bm{E}}}
\newcommand{\bH}{{\bm{H}}}
\newcommand{\bp}{{\bm{p}}}
\newcommand{\bP}{{\bm{P}}}
\newcommand{\bq}{{\bm{q}}}
\newcommand{\bQ}{{\bm{Q}}}

\newcommand{\rA}{{\rm A}}
\newcommand{\rB}{{\rm B}}

\newcommand{\Pc}{{P_{\rm c}}}

\newcommand{\Tl}{{TlCuCl$_3$}}
\newcommand{\K}{{KCuCl$_3$}}
\newcommand{\NH}{{NH$_4$CuCl$_3$}}
\newcommand{\CsNi}{{CsNiCl$_3$}}
\newcommand{\CsFe}{{CsFeCl$_3$}}
\newcommand{\Ba}{{BaCo$_2$V$_2$O$_8$}}
\newcommand{\Ca}{{Ca$_2$RuO$_4$}}
\newcommand{\DLCB}{{C$_9$H$_{18}$N$_2$CuBr$_4$}}

\begin{document}

\title{
First ESR Detection of Higgs Amplitude Mode and Analysis with Extended Spin-Wave Theory in Dimer System \K
}

\author{Masashige Matsumoto}
\email{matsumoto.masashige@shizuoka.ac.jp}
\affiliation{Department of Physics, Faculty of Science, Shizuoka University, Shizuoka 422-8529, Japan}

\author{Takahiro Sakurai}
\affiliation{Research Facility Center for Science and Technology, Kobe University, Kobe 657-8501, Japan}

\author{Yuki Hirao}
\affiliation{Graduate School of Science, Kobe University, Kobe 657-8501, Japan}

\author{Hitoshi Ohta}
\affiliation{Molecular Photoscience Research Center, Kobe University, Kobe 657-8051, Japan}

\author{Yoshiya Uwatoko}
\affiliation{Institute for Solid State Physics, University of Tokyo, Chiba 277-8581, Japan}

\author{Hidekazu Tanaka}
\affiliation{Department of Physics, Tokyo Institute of Technology, Tokyo 152-8551, Japan}

\date{Received: 15 September 2020 / Revised: 5 November 2020 / Accepted: 9 November 2020}

%%%%%%%%%%%%%%%%%%%%%%%%%%%%%%%%%%%%%%%%%%%%%%%%%%%%%%%%%%%%%%%%%%%%%%%%
\begin{abstract}
\K~is known to show a quantum phase transition from the disordered to antiferromagnetically ordered phases by applying pressure.
There is a longitudinal excitation mode (Higgs amplitude mode) in the vicinity of the quantum critical point in the ordered phase.
To detect the Higgs amplitude mode, high-pressure ESR measurements are performed in \K.
The experimental data are analyzed by the extended spin-wave theory on the basis of the vector spin chirality.
We report the first ESR detection of the Higgs amplitude mode
and the important role of the electric dipole described by the vector spin chirality.
\end{abstract}

\maketitle

%%%%%%%%%%%%%%%%%%%%%%%%%%%%%%%%%%%%%%%%%%%%%%%%%%%%%%%%%%%%%%%%%%%%%%%%
\section{Introduction}
%%%%%%%%%%%%%%%%%%%%%%%%%%%%%%%%%%%%%%%%%%%%%%%%%%%%%%%%%%%%%%%%%%%%%%%%

After the discovery of the Higgs boson in high-energy physics,
searching for similar excitations in condensed matters attracted interests
\cite{Pekker-2015}.
In the broken-symmetry phase, the collective excitations can be characterized
by fluctuations of phase and amplitude of the order parameter.
The phase fluctuation corresponds to the massless Nambu--Goldstone mode,
whereas the amplitude fluctuation has similarity to the massive Higgs boson,
and the excitation was recently termed as ``Higgs amplitude mode" in condensed matter physics
\cite{Pekker-2015}.
The Higgs amplitude mode spontaneously decays into a pair of low-energy Nambu--Goldstone modes
and this disturbs its observation
\cite{Chubukov-1994,Sachdev-1999-1,Sachdev-1999-2}.
The effect becomes prominent in low dimensional systems or for large energy gap of the Higgs amplitude mode,
where the latter means that the order parameter is substantially developed.
Therefore, three-dimensionally interacting system in the vicinity of a quantum critical point is favorable
for observing the Higgs amplitude mode.

Spin-wave is an elementary excitation in magnets accompanied by fluctuations of a magnetic moment.
There are two characteristic directions; one is perpendicular to the moment and the other is along the moment.
The perpendicular fluctuation corresponds to the transverse (Nambu--Goldstone) mode,
whereas the longitudinal one corresponds to the longitudinal (Higgs amplitude) mode.
In conventional magnets, the ordered moment is substantial and hard to modulate its amplitude.
The low-energy excitations are formed by the Nambu--Goldstone mode,
whereas the Higgs amplitude mode is difficult to detect in conventional magnets.

In quantum magnets, the Higgs amplitude mode has been observed, for instance, in
\CsNi~\cite{Morra-1988,Tun-1990,Kakurai-1991},
\Tl~\cite{Ruegg-2008,Kuroe-2012,Matsumoto-2008,Merchant-2014},
\Ba~\cite{Grenier-2015},
\Ca~\cite{Jain-2017},
and \DLCB~\cite{Hong-2017,Ying-2019}.
For the triangular antiferromagnet \CsNi, the longitudinal and transverse fluctuations are hybridized
due to the noncollinearity of the stabilized 120$^\circ$ structure
\cite{Affleck-1992},
and it was recently found in \CsFe~that the hybridization plays an important role
in the evolution of the spin-wave excitations under the pressure through the quantum critical point
\cite{Hayashida-2019,Matsumoto-2020}.
To have pure Higgs amplitude mode (longitudinal mode), collinear magnetic structure is required.

Among the above examples, \Tl~and its isostructural compound \K~are ideal systems to observe the Higgs amplitude mode.
They are three-dimensionally interacting spin dimer systems
and are known to have a disordered ground state with a finite excitation gap
\cite{Takatsu-1997,Shiramura-1997}.
Applying external magnetic field can collapse the gap, and the field-induced order appears above the critical field
\cite{Oosawa-1999,Tanaka-2001,Oosawa-2002},
which is interpreted as a magnon Bose--Einstein condensation (BEC)
\cite{Nikuni-2000}.
Another way to stabilize the ordered phase is to apply pressure
\cite{Ruegg-2008,Tanaka-2003,Oosawa-2003,Oosawa-2004,Goto-2004,Ruegg-2004,Goto-2006,Goto-2007},
where a collinear antiferromagnetic structure is stabilized.
In these compounds, the critical pressure is less than 1 GPa, and it is reachable by experiments.
In the pressure-induced ordered phase, the pure Higgs amplitude mode was expected
\cite{Chubukov-1995,Normand-1997,Matsumoto-2004}
and was actually observed in \Tl~\cite{Ruegg-2008,Kuroe-2012}.
On the other hand, there is no work reporting the Higgs amplitude mode in the isostructural compound \K.
Besides, there is no report observing the Higgs amplitude mode by ESR.

Thus far, inelastic neutron scattering \cite{Ruegg-2008} and Raman scattering  \cite{Kuroe-2012,Matsumoto-2008}
are the possible measurements for the Higgs amplitude mode in \Tl.
In the present work, we try to detect the Higgs amplitude mode by high-pressure ESR.
The high-pressure ESR is a very powerful means to know the spin states of  magnetic materials under pressure.
In particular, the recently developed our broadband high-pressure ESR system in the THz region
has features of high pressure up to 2.5 GPa and a wide frequency range up to 0.8 THz
\cite{Sakurai-2015,Sakurai-2018-1}.
These features are very useful for observing pressure changes in the ground state
and the low-lying excited states of quantum magnets,
because the THz region matches their energy levels of the low-lying excited states.
For example, we observed the pressure-induced quantum phase transition from the dimer singlet state
to the plaquette singlet state in the two-dimensional orthogonal dimer spin system SrCu$_2$(BO$_3$)$_2$
\cite{Sakurai-2018-2},
the novel magnetic phases under pressure and high magnetic field in the triangular lattice antiferromagnet Cs$_2$CuCl$_4$
\cite{Zvyagin-2019},
anomalies in the antiferromagnetic resonance mode corresponding to the 1/3 magnetization plateau
in the hexagonal ABX$_3$-type antiferromagnet CsCuCl$_3$
\cite{Okubo-2019},
and so on.
On the other hand, \K, which is of interest in this paper,
has not been studied sufficiently around the critical pressure from the microscopic point of view, as mentioned above.
Furthermore, the ESR transition from the ground state to the first excited state
with the energy gap around 0.6 THz is in our measurable region
and the critical pressure of 0.82 GPa is also easily accessible.
Therefore, we tried the high-pressure ESR measurement in detail
to observe the pressure dependence of the gap energy and the Higgs amplitude mode.

In spin dimer systems, the direct transition from the singlet to triplet states has been observed
by ESR in \K~and \Tl~\cite{Tanaka-1998,Kimura-2004}.
In the conventional interpretation, however, the direct transition is forbidden.
Recently, it was pointed out that an electric dipole described by vector spin chirality is the origin of the direct transition
\cite{Kimura-2018,Kimura-2020}.
Along this line, we examine the observed ESR intensity.

This paper is organized as follows.
In Sect. 2, magnon dispersion relation in the disordered phase is summarized.
The formula for the dispersion relation is used to analyze the pressure dependence of the exchange interactions in \K.
In Sect. 3, we derive ESR intensity in the disordered phase on the basis of the vector spin chirality.
We present experimental results in Sect. 4.
In Sect. 5, we study ESR intensity in the ordered phase and analyze the data to show that the Higgs amplitude mode is detected by ESR.
The last section gives a summary of this paper.
In the Appendix, we give details of the extended spin-wave theory, or the generalized Holstein--Primakoff theory,
for the magnon dispersion in the ordered phase
\cite{Papanicolaou-1984,Onufrieva-1985,Joshi-1999,Shiina-2003,Shiina-2004}.
It is used to know how the Higgs amplitude mode evolves with the magnetic field.

%%%%%%%%%%%%%%%%%%%%%%%%%%%%%%%%%%%%%%%%%%%%%%%%%%%%%%%%%%%%%%%%%%%%%%%%
\section{Magnon Dispersion Relation in Disordered Phase}
%%%%%%%%%%%%%%%%%%%%%%%%%%%%%%%%%%%%%%%%%%%%%%%%%%%%%%%%%%%%%%%%%%%%%%%%

%%%%%%%%%%%%%%%%%%%%%%%%%%%%%%%%%%%%%%%%%%%%%%%%%%%%%%%%%%%%%%%%%%%%%%%%
\begin{figure}[t]
\begin{center}
\includegraphics[width=6cm,clip]{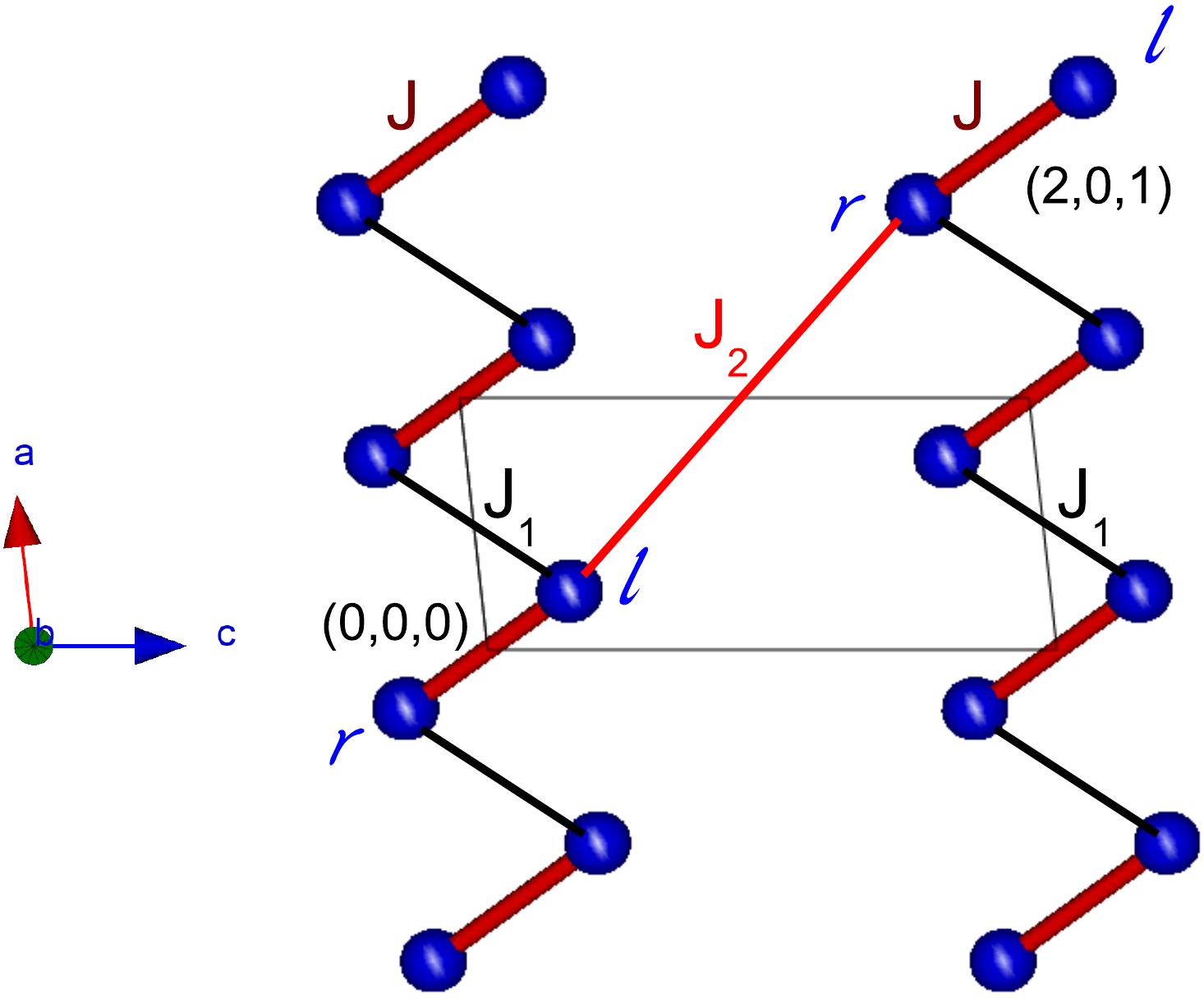}
\end{center}
\begin{center}
\includegraphics[width=7.5cm,clip]{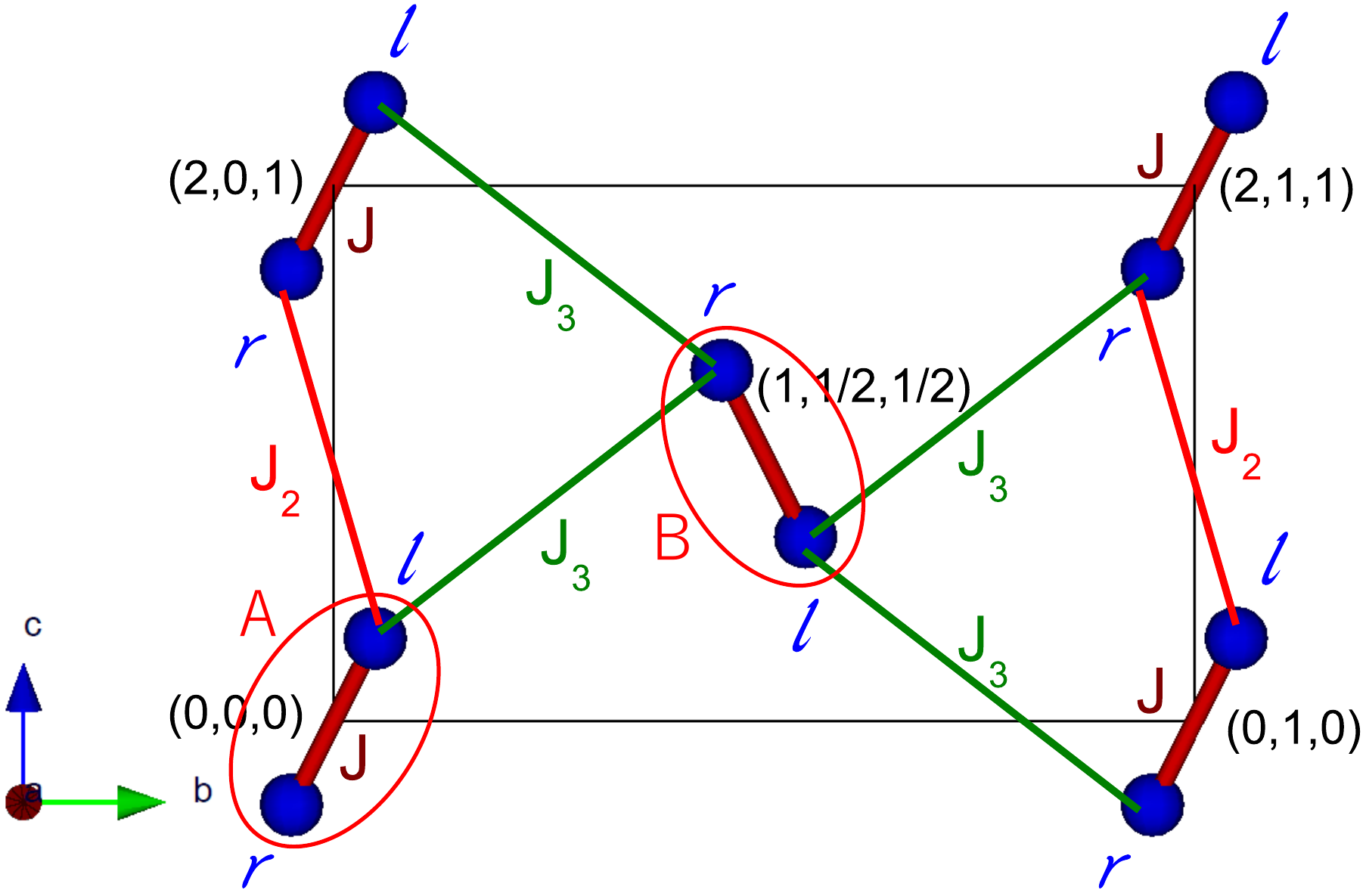}
\caption{
(Color online)
Schematics of the crystal structure of \K.
$l$ and $r$ represent the left and right side of a dimer, respectively.
The position of each dimer is depicted.
$J$ represents the intradimer interaction, whereas $J_1$, $J_2$ and $J_3$ are interdimer interactions.
The four Heisenberg interactions are antiferromagnetic and they are expressed with positive values.
Dimer sites at the corner and the center in the unit cell are described as A and B sublattices, respectively.
}
\label{fig:interaction}
\end{center}
\end{figure}
%%%%%%%%%%%%%%%%%%%%%%%%%%%%%%%%%%%%%%%%%%%%%%%%%%%%%%%%%%%%%%%%%%%%%%%%

The schematic of the crystal structure of \K~is shown in Fig. \ref{fig:interaction}.
The solid red line represents an intradimer interaction.
There are two dimer sites in the unit cell.
Dimers at the corner and the center are described as A and B sublattices, respectively.

In the disordered phase, magnon dispersion relation is analytically expressed on the basis of a random phase approximation
\cite{Cavadini-1999,Cavadini-2001,Oosawa-PRB-2002}
and a harmonic bond-operator formulation
\cite{Matsumoto-2004,Sommer-2001,Matsumoto-2002}.
Both theories lead to the same formula.
Here, we briefly introduce the bond-operator formulation.
For a dimer, we define the following creation Bose operators for triplet states
\cite{Sachdev-1990}
\begin{align}
\begin{aligned}
&t_+^\dagger |{\rm vac}\rangle = |\uparrow\uparrow\rangle, \cr
&t_0^\dagger |{\rm vac}\rangle = \frac{1}{\sqrt{2}} ( |\uparrow\downarrow\rangle + |\downarrow\uparrow\rangle ), \\
&t_-^\dagger |{\rm vac}\rangle = |\downarrow\downarrow\rangle.
\end{aligned}
\end{align}
Here, $|{\rm vac}\rangle$ represents the vacuum state.
Under a singlet ground state in the disordered phase, the Hamiltonian for the triplet excitations are written as
\cite{Normand-1997,Matsumoto-2004,Matsumoto-2002,Gopalan-1994}
\begin{align}
\H &= \sum_\bk \left[ \sum_{\alpha=+,0,-} ( \epsilon_\bk - \alpha g\mu_{\rm B} H ) t_{\bk\alpha}^\dagger t_{\bk\alpha} \right. \cr
&~~~~~~~~~~~~\left.
+ \frac{1}{2} \gamma_\bk \left( t_{-\bk 0}^\dagger t_{\bk 0}^\dagger + t_{\bk 0} t_{-\bk 0} \right) \right. \cr
&~~~~~~~~~~~~\left.
- \frac{1}{2} \gamma_\bk \left( t_{-\bk -}^\dagger t_{\bk +}^\dagger + t_{\bk +} t_{-\bk -} \right) \right. \cr
&~~~~~~~~~~~~\left.
- \frac{1}{2} \gamma_\bk \left( t_{-\bk +}^\dagger t_{\bk -}^\dagger + t_{\bk -} t_{-\bk +} \right) \right].
\label{eqn:H-ana}
\end{align}
Here, $g$ is a $g$-factor, $\mu_{\rm B}$ is the Bohr magneton, and $H$ is an external magnetic field.
We introduced the same Bose operators on the A and B sublattices,
and expressed the Hamiltonian in the expanded scheme for the Brillouin zone.
$t_{\bk \alpha}$ is the Fourier transformed operator defined by
\begin{align}
t_{\bk \alpha} = \frac{1}{\sqrt{N}} \sum_i e^{-i\bk\cdot\br_i} t_{i \alpha},
\end{align}
where $\br_i$ represents the position of the $i$th dimer site and $N$ is number of the dimers in the sample.
In Eq. (\ref{eqn:H-ana}), $\epsilon_\bk$ and $\gamma_\bk$ are defined by
\begin{align}
\begin{aligned}
\epsilon_\bk &= J + \gamma_\bk, \cr
\gamma_\bk &= - \frac{1}{2} J_1 \cos{k_a} - \frac{1}{2} J_2 \cos(2k_a+k_c) \cr
&~~~
- J_3 \cos\left(k_a+\frac{k_c}{2}\right)\cos{\frac{k_b}{2}}.
\label{eqn:gamma}                           
\end{aligned}
\end{align}
Here, $k_a$, $k_b$, and $k_c$ are wavenumber along the $a^*$, $b^*$, and $c^*$ directions in the reciprocal lattice space, respectively.
The exchange interactions ($J$, $J_1$, $J_2$, $J_3$) are depicted in Fig. \ref{fig:interaction}.

The Hamiltonian in Eq. (\ref{eqn:H-ana}) can be diagonalized by the following Bogoliubov transformation:
\begin{align}
&t_{\bk 0} = u_\bk \alpha_{\bk 0} - v_\bk \alpha_{-\bk 0}^\dagger, \cr
&t_{\bk +} = u_\bk \alpha_{\bk +} + v_\bk \alpha_{-\bk -}^\dagger, \label{eqn:trans} \\
&t_{\bk -} = u_\bk \alpha_{\bk -} + v_\bk \alpha_{-\bk +}^\dagger.
\nonumber
\end{align}
Here, $u_\bk$ and $v_\bk$ are coefficients for the transformation and are real numbers.
The $\alpha_{\bk m}$ ($m=1,0,-1)$ operators describes the spin-wave excitations.
To satisfy the bosonic commutation relation for $\alpha_{\bk m}$,
$u_\bk$ and $v_\bk$ are subjected to fulfill $u_\bk^2 - v_\bk^2 = 1$.
They are given by
\begin{align}
\begin{aligned}
&u_\bk = \sqrt{ \frac{1}{2} \left( \frac{\epsilon_\bk}{E_\bk} + 1 \right) }, \cr
&v_\bk = \sqrt{ \frac{1}{2} \left( \frac{\epsilon_\bk}{E_\bk} - 1 \right) } \frac{\gamma_\bk}{|\gamma_\bk|}.
\end{aligned}
\end{align}
Substituting Eq. (\ref{eqn:trans}) into Eq. (\ref{eqn:H-ana}), we obtain the following Hamiltonian in a diagonal form:
\begin{align}
\H = \sum_\bk \sum_{m=+,0,-} E_{\bk m} \alpha_{\bk m}^\dagger \alpha_{\bk m},
\label{eqn:H-diag}
\end{align}
where
\begin{equation}
\begin{aligned}
&E_{\bk m}
= E_\bk - g\mu_{\rm B} H m~~~~~~(m=1,0,-1), \cr
&E_\bk
= \sqrt{ \epsilon_\bk^2 - \gamma_\bk^2}
= \sqrt{ J^{2} + 2J \gamma_\bk }.
\label{eqn:E-k}
\end{aligned}
\end{equation}
In Eq. (\ref{eqn:H-diag}), we dropped a constant term.
$E_{\bk m}$ represents dispersion relation of the triplet excitation under an external magnetic field,
whereas $E_\bk$ is that at $H=0$.
Notice that there are three excitation modes in the extended zone scheme.
The formula of Eq. (\ref{eqn:E-k}) was used to fit the data
obtained by inelastic neutron scattering measurements in the disordered phase
\cite{Matsumoto-2004,Cavadini-1999,Matsumoto-2002}.
The estimated values of the exchange interaction parameters are summarized in Table \ref{table:J}.

%%%%%%%%%%%%%%%%%%%%%%%%%%%%%%%%%%%%%%%%%%%%%%%%%%%%%%%%%%%%%%%%%%%%%%%%
\begin{table}
\caption{
Exchange interaction parameters (THz) at the ambient pressure (0 GPa) for \K.
They were extracted from the observed magnon dispersion relation with Eq. (\ref{eqn:E-k})
\cite{Cavadini-1999,Cavadini-2001}.
Parameters for \Tl~are also shown here for comparison.
}
\label{table:J}
\begin{center}
\begin{tabular}{ccccc}
\hline
Material & $J(0)$ & $J_1(0)$ & $J_2(0)$ & $J_3(0)$ \\
\hline
\K  & 1.026 & 0.101 & 0.191 & 0.176 \\
\hline
\Tl & 1.330 & 0.104 & 0.764 & 0.220 \\
\hline      
\end{tabular}
\end{center}
\end{table}
%%%%%%%%%%%%%%%%%%%%%%%%%%%%%%%%%%%%%%%%%%%%%%%%%%%%%%%%%%%%%%%%%%%%%%%%

Since we take the left and right side of dimers as shown in Fig. \ref{fig:interaction},
the excitation gap (minimum excitation energy) is located at $\bk=\bQ=(0,0,0)$,
at which $\epsilon_\bk$ takes the minimum value.
Here, $\bQ$ is an order wavevector characterizing the magnetic structure in the ordered phase
\cite{note:Q}.
As in Eq. (\ref{eqn:E-k}), applying a magnetic field can reduce the excitation gap, and this leads to a field-induced magnetic order.
Another way to stabilize the ordered phase is to modify the exchange interactions.
This is realized by applying pressure and a pressure-induced order takes place
\cite{Ruegg-2008,Tanaka-2003,Oosawa-2003,Oosawa-2004,Goto-2004,Ruegg-2004,Goto-2006,Goto-2007}.
For $H=0$, the excitation gap is expressed as
\begin{align}
E_{\rm gap}
= \sqrt{ J^{2} + 2J \gamma_\bQ }
= \sqrt{ J[J - (J_1+J_2+2J_3)]}.
\label{eqn:e-gap}
\end{align}
The enhancement of $J$ increases the gap, while that of $J_1+J_2+2J_3$ decreases the gap.
There is a competition between the intradimer and interdimer interactions.
The former favors the disordered phase, whereas the latter favors the ordered phase.
By measuring magnetization of \K, it is reported that $J$ is reduced and $J_1+J_2+2J_3$ is enhanced by applying pressure
\cite{Goto-2006}.
Thus, there is a quantum critical point at which the disordered and ordered phases are separated.
The excitation gap decreases with pressure and becomes soft at a critical pressure,
maintaining the threefold degeneracy.
Above the critical pressure, a finite antiferromagnetic moment appears.
The excitation modes then split into twofold degenerate transverse (Nambu--Goldstone) modes
and a single longitudinal (Higgs amplitude) mode
\cite{Matsumoto-2004}.
When the pressure is increased further,
the excitation energy of the Nambu--Goldstone mode stays zero energy at $\bk=\bQ$,
whereas the Higgs amplitude mode acquires an energy gap which evolves with the pressure.
\cite{Matsumoto-2004}.
These behaviors were actually observed in \Tl~by inelastic neutron scattering under pressures,
and the Higgs amplitude mode was confirmed
\cite{Ruegg-2008}.

ESR is another possible measurement to detect the Higgs amplitude mode.
In ESR, an external magnetic field is required to cause a resonance
under the irradiation of a microwave with a fixed frequency.
Thus far, there is no report observing the Higgs amplitude mode by ESR.
Since energies of excitation modes are not easy to represent in analytic forms under a finite field in the ordered phase,
we use the extended spin-wave theory to analyze the field dependence of the excitation modes and the ESR intensity under pressures.
Details of the formulation are given in the Appendix.

%%%%%%%%%%%%%%%%%%%%%%%%%%%%%%%%%%%%%%%%%%%%%%%%%%%%%%%%%%%%%%%%%%%%%%%%
\section{Intensity of ESR in Disordered Phase}
%%%%%%%%%%%%%%%%%%%%%%%%%%%%%%%%%%%%%%%%%%%%%%%%%%%%%%%%%%%%%%%%%%%%%%%%

Recent studies on ESR in spin dimer systems revealed that both magnetic and electric channels
play important roles in the optical transition process
\cite{Kimura-2018,Kimura-2020}.
In the disordered phase, the magnetic transition is forbidden and the electric channel becomes important.
In this section, we present the formulation and give matrix elements for the transition,
with which the electric transition is completely described in the disordered phase.
It is helpful to understand ESR intensity also in the ordered phase and useful for future works.

%%%%%%%%%%%%%%%%%%%%%%%%%%%%%%%%%%%%%%%%%%%%%%%%%%%%%%%%%%%%%%%%%%%%%%%%
\subsection{Magnetic Channel}
%%%%%%%%%%%%%%%%%%%%%%%%%%%%%%%%%%%%%%%%%%%%%%%%%%%%%%%%%%%%%%%%%%%%%%%%

In the conventional ESR, the uniform component of spin operators in a dimer, $\bS_l+\bS_r$, concerns the transition (magnetic channel).
The matrix forms of the operators in the singlet and triplet basis
($|00\rangle, |11\rangle, |10\rangle, |1-1\rangle$)
are given by
\begin{align}
&S_l^x + S_r^x \longrightarrow \frac{1}{\sqrt{2}}
\begin{pmatrix}
0 & 0 & 0 & 0 \cr
0 & 0 & 1 & 0 \cr
0 & 1 & 0 & 1 \cr
0 & 0 & 1 & 0
\end{pmatrix}, \cr
&S_l^y + S_r^y \longrightarrow \frac{1}{\sqrt{2}}
\begin{pmatrix}
0 & 0 & 0 & 0 \cr
0 & 0 & -i & 0 \cr
0 & i & 0 & -i \cr
0 & 0 & i & 0
\end{pmatrix}, \label{eqn:S+} \\
&S_l^z + S_r^z \longrightarrow
\begin{pmatrix}
0 & 0 & 0 & 0 \cr
0 & 1 & 0 & 0 \cr
0 & 0 & 0 & 0 \cr
0 & 0 & 0 & 1
\end{pmatrix}.
\nonumber
\end{align}
We can see that there is no matrix element between the singlet and triplet states.
This indicates that the direct (singlet--triplet) transition is forbidden in the conventional ESR.
Note that this feature also holds in the presence of an XYZ type anisotropy in the intradimer interaction.
In the presence of a Dzyaloshinskii--Moriya (DM) interaction, on the other hand,
a finite matrix elements appears between the ground and excited states
and this enables the measurement of ESR in the disordered phase
\cite{Matsumoto-2008-DM,Sakai-2000,Sakai-2003}.
In \K~and \Tl, however, there is an inversion center in a dimer and this excludes the DM interaction.
Thus, the direct transition is difficult to understand by the conventional magnetic channel.

%%%%%%%%%%%%%%%%%%%%%%%%%%%%%%%%%%%%%%%%%%%%%%%%%%%%%%%%%%%%%%%%%%%%%%%%
\subsection{Electric Channel}
%%%%%%%%%%%%%%%%%%%%%%%%%%%%%%%%%%%%%%%%%%%%%%%%%%%%%%%%%%%%%%%%%%%%%%%%

%%%%%%%%%%%%%%%%%%%%%%%%%%%%%%%%%%%%%%%%%%%%%%%%%%%%%%%%%%%%%%%%%%%%%%%%
\subsubsection{Vector Spin Chirality}
%%%%%%%%%%%%%%%%%%%%%%%%%%%%%%%%%%%%%%%%%%%%%%%%%%%%%%%%%%%%%%%%%%%%%%%%

The vector spin chirality of a dimer at the $i$th cite is defined by
\begin{align}
\bW_i = \bS_{il} \times \bS_{ir}.
\label{eqn:vector-spin}
\end{align}
This seems similar to the DM interaction.
It is known that the DM interaction can present in the absence of the inversion center between the two spins.
On the other hand, the vector spin chirality can survive even in the presence of the inversion center and describes an electric dipole
\cite{Kaplan-2011}.
This is owing to the fact that the vector spin chirality is symmetric with respect to the time-reversal operation
and has odd parity for the space inversion.
Notice that these features are the same as those of the electric dipole.
The difference between the DM interaction and the vector spin chirality is that
the former is a Hamiltonian, while the latter is an operator.
In a spin dimer, it was proposed that the spin-dependent electric dipole is described by the vector spin chirality
\cite{Kimura-2018,Kimura-2016,Kimura-2017}.
This was confirmed by observing ferroelectricity whose magnitude is proportional to the expectation value of the vector spin chirality
in the magnon BEC phase of \Tl~\cite{Kimura-2016}.

At each dimer site, matrix forms of the vector spin chirality are given by the singlet and triplet basis as
\begin{align}
&W^x= S_l^y S_r^z - S_l^z S_r^y \longrightarrow \frac{1}{2\sqrt{2}}
\begin{pmatrix}
0 & -i & 0 & i \cr
i  & 0 & 0 & 0 \cr
0 & 0 & 0 & 0 \cr
-i & 0 & 0 & 0
\end{pmatrix}, \cr
&W^y= S_l^z S_r^x - S_l^x S_r^z \longrightarrow \frac{1}{2\sqrt{2}}
\begin{pmatrix}
0 & 1 & 0 & 1 \cr
1 & 0 & 0 & 0 \cr
0 & 0 & 0 & 0 \cr
1 & 0 & 0 & 0
\end{pmatrix}, \label{eqn:v-matrix} \\
&W^z= S_l^x S_r^y - S_l^y S_r^x \longrightarrow \frac{1}{2}
\begin{pmatrix}
0 & 0 & i & 0 \cr
0 & 0 & 0 & 0 \cr
-i & 0 & 0 & 0 \cr
0 & 0 & 0 & 0
\end{pmatrix}.
\nonumber
\end{align}
The vector spin chirality has finite matrix elements between the singlet and triplet states,
and the direct transition is possible by the electric field component of the microwave (electric channel)
\cite{Kimura-2018,Kimura-2020,Kimura-2016,Kimura-2017},
differing from the magnetic channel shown in Eq. (\ref{eqn:S+}).
This is related to the fact that the singlet and triplet states have odd and even parities, respectively.
The magnetic dipole $\bS_l+\bS_r$ has an even parity and it cannot connect the singlet and triplet states.
On the other hand, the vector spin chirality $\bS_l\times\bS_r$ has an odd parity and has the finite matrix elements.

%%%%%%%%%%%%%%%%%%%%%%%%%%%%%%%%%%%%%%%%%%%%%%%%%%%%%%%%%%%%%%%%%%%%%%%%
\begin{figure}
\begin{center}
\includegraphics[width=5.5cm]{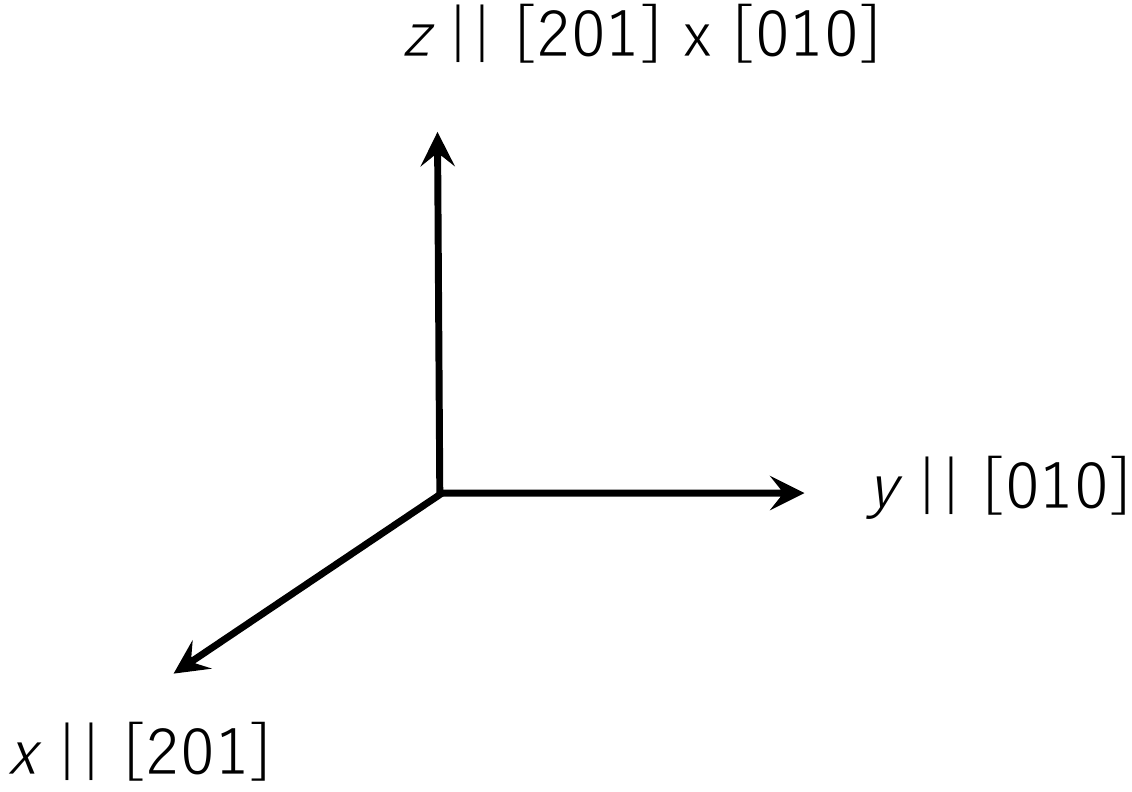}
\caption{
The relation between the crystal axes and $xyz$ coordinates
\cite{Kimura-2020}.
Notice that the $z$-axis is perpendicular to the $(10\bar{2})$ plane [$z\perp (10\bar{2})$].
In the present ESR measurements, the external magnetic field is applied in the $x$ direction with the Faraday configuration.
The alternating electromagnetic fields ($\bE^\omega,\bH^\omega)$ are in the $yz$-plane.
}
\label{fig:coordinate}
\end{center}
\end{figure}
%%%%%%%%%%%%%%%%%%%%%%%%%%%%%%%%%%%%%%%%%%%%%%%%%%%%%%%%%%%%%%%%%%%%%%%%

Since $\bS_l\times\bS_r$ is parity odd, it can couple to the electric field and plays a role of an operator for the electric dipole.
In general, an electric dipole is expressed as
\cite{Kaplan-2011}
\begin{align}
p_i^\alpha = \sum_{\beta=x,y,z} C^\alpha_\beta(i) W_i^\beta.~~~(\alpha=x,y,z)
\label{eqn:p}
\end{align}
Here, $p_i^\alpha$ is the $\alpha$ component of the electric dipole at the $i$th site.
$W_i^\beta$ represents the $\beta$ component of $\bW_i$.
$C^\alpha_\beta(i)$ is a coefficient tensor, which depends on the sublattice ($i={\rm A},{\rm B}$).
Equation (\ref{eqn:p}) is a generalization of a spin-driven electric dipole by the inverse DM effect
\cite{Katsura-2005,Mostovoy-2006,Sergienko-2006,Jia-2007}.
The electric dipole in Eq. (\ref{eqn:p}) is not induced by an external electric field.
It is an operator and can present in the absence of the electric field.
In the presence of an external electric field, the dipole operator linearly couples to the electric field.

%Following Ref. \ref{ref:Kimura-2020}, we introduce $xyz$ coordinates for \K~(see Fig. \ref{fig:coordinate}).
Following Ref. 43, we introduce $xyz$ coordinates for \K~(see Fig. \ref{fig:coordinate}).
The $x$-axis is taken along the [201] direction.
The $y$-axis is taken along the [010] direction.
The $z$-axis is then taken along the $[201]\times[010]$ direction, which is perpendicular to the $(10\bar{2})$ plane.
The dimers A and B are related to the symmetry transformation by a 2${_1}$ symmetry operation along the [010] direction ($y$-axis).
The coefficient tensors $C(i)$ for the A and B dimer are related as
$C({\rm B})=R_y(\pi) C({\rm A}) R_y^{-1}(\pi)$,
where $R_y(\pi)$ is a matrix for $\pi$ rotation around the $y$-axis
\cite{Kimura-2018,Kimura-2020,Kimura-2017}.
We obtain the following expressions:
\begin{align}
\begin{aligned}
&C({\rm A})=
\begin{bmatrix}
C^x_x & C^x_y & C^x_z \\
C^y_x & C^y_y & C^y_z \\
C^z_x & C^z_y & C^z_z
\end{bmatrix}, \cr
&C({\rm B})=
\begin{bmatrix}
C^x_x & -C^x_y & C^x_z \\
-C^y_x & C^y_y & -C^y_z \\
C^z_x & -C^z_y & C^z_z
\end{bmatrix}.
\label{eqn:C-AB}
\end{aligned}
\end{align}
The electric dipole in the $i$th cite is then written as
\begin{align}
p_i^\alpha
&= \left[ C^\alpha_\beta(+) + C^\alpha_\beta(-) e^{-i\bk_\rB\cdot\br_i} \right] W_i^\beta.
\label{eqn:p_i}
\end{align}
For the later convenience, we put $\bk_\rB=(0,0,2\pi)$
and introduced the following uniform and staggered components of the coefficient tensor:
\begin{equation}
\begin{aligned}
&C(+) = \frac{1}{2} [C(\rA) + C(\rB) ] =
\begin{bmatrix}
C^x_x & 0 & C^x_z \\
0 & C^y_y & 0 \\
C^z_x & 0 & C^z_z
\end{bmatrix}, \cr
&C(-) = \frac{1}{2} [C(\rA) - C(\rB) ] =
\begin{bmatrix}
0 & C^x_y & 0 \\
C^y_x & 0 & C^y_z \\
0 & C^z_y & 0
\end{bmatrix}.
\end{aligned}
\label{eqn:C-pm}
\end{equation}
In Eq. (\ref{eqn:p_i}), notice that $e^{-i\bk_\rB\cdot\br_i}=\pm 1$ on the A and B sublattices, respectively.

There are nine free coefficients.
In the presence of various symmetries in a dimer, such as rotation and mirror symmetries,
the number of free coefficients is reduced.
In this case, the coefficient tensor was classified by the symmetries
\cite{Kaplan-2011}.
For the spin dimer in \K~and \Tl, there is no symmetry except for the space inversion and the nine coefficients remain.
In the presence of the inversion symmetry, even and odd parities are distinguishable
and the electric dipole is described by the antisymmetric vector spin chirality.
When there is no inversion center in a dimer, even and odd parities are not distinguishable.
The electric dipole is then also described by symmetric operators such as $S_l^\alpha S_r^\beta + S_l^\beta S_r^\alpha$,
where the coefficient tensor for this was classified by the symmetries in a dimer
\cite{Matsumoto-2017}.

%%%%%%%%%%%%%%%%%%%%%%%%%%%%%%%%%%%%%%%%%%%%%%%%%%%%%%%%%%%%%%%%%%%%%%%%
\subsubsection{Transition Probability}
%%%%%%%%%%%%%%%%%%%%%%%%%%%%%%%%%%%%%%%%%%%%%%%%%%%%%%%%%%%%%%%%%%%%%%%%

The selection rule for ESR in the disordered phase was investigated in \K~on the basis of the vector spin chirality with the coefficient tensor
\cite{Kimura-2018}.
We proceed with the investigation and derive ESR intensity for various field directions.

We assume that a microwave is irradiated to the sample under a uniform static magnetic field applied in the $z$ direction.
The time-dependent perturbation Hamiltonian is expressed as
\begin{align}
\begin{aligned}
&\H'(t) = \H'_P e^{-i\omega t}, \cr
&\H'_P = - \bE^\omega \cdot \bP, \cr
&\bP=\sum_i \bp_i.
\label{eqn:Hp}
\end{aligned}
\end{align}
Here, $\bE^\omega$ is for an alternating electric field with an angular frequency $\omega$.
$\bP$ is the total electric dipole.
At low temperatures, the transition probability from the ground state (GS) to an excited state (ES) is given by the Fermi's Golden rule as
\begin{align}
W
&= 2\pi \left| \langle {\rm ES}| \H'_P | {\rm GS} \rangle \right|^2 \delta( E_{\rm ES} - \omega) \cr
&=2\pi \left| \langle {\rm ES}| ( E_x^\omega P^x + E_y^\omega P^y + E_z^\omega P^z ) | {\rm GS} \rangle \right|^2
       \delta( E_{\rm ES} - \omega). \cr
\label{eqn:W}
\end{align}
Here, $E_{\rm ES}$ represents the energy of the excited state.
Using Eq. (\ref{eqn:p_i}), we can express the total electric dipole as
\begin{align}
P^\alpha
&= \sum_i \left[ C^\alpha_\beta(+) + C^\alpha_\beta(-) e^{-i\bk_\rB\cdot\br_i} \right] W_i^\beta \cr
&= \sqrt{N} \left[ C^\alpha_\beta(+) W^\beta(0) + C^\alpha_\beta(-) W^\beta(\bk_\rB) \right],
\label{eqn:P0}
\end{align}
where we introduced the following Fourier transformation for $\bW$:
\begin{align}
\bW(\bq) = \frac{1}{\sqrt{N}} \sum_i e^{-i\bq\cdot\br_i} \bW_i.
\label{eqn:W-Fourier}
\end{align}
According to the matrix form of the vector spin chirality in Eq. (\ref{eqn:v-matrix}),
the operators at the $i$th site are expressed as
\begin{align}
&W_i^x = \frac{i}{2\sqrt{2}} ( t_{i+}^\dagger - t_{i-}^\dagger - t_{i+} + t_{i-} ), \cr
&W_i^y = \frac{1}{2\sqrt{2}} ( t_{i+}^\dagger + t_{i-}^\dagger + t_{i+} + t_{i-} ), \label{eqn:wi} \\
&W_i^z = - \frac{i}{2} ( t_{i0}^\dagger - t_{i0} ).
\nonumber
\end{align}
Here, we replaced the Bose operator for the singlet state as a c-number 1.
Using Eq. (\ref{eqn:W-Fourier}), we obtain
\begin{align}
&W^x(\bq) = \frac{i}{2\sqrt{2}} ( t_{-\bq+}^\dagger - t_{-\bq-}^\dagger - t_{\bq+} + t_{\bq-} ), \cr
&W^y(\bq) = \frac{1}{2\sqrt{2}} ( t_{-\bq+}^\dagger + t_{-\bq-}^\dagger + t_{\bq+} + t_{\bq-} ), \label{eqn:O2} \\
&W^z(\bq) = - \frac{i}{2} ( t_{-\bq0}^\dagger - t_{\bq0} ).
\nonumber
\end{align}
The triplet operators are expressed by the operators $\alpha_{\bk m}$ for the spin-wave excitation
via the Bogoliubov transformation in Eq. (\ref{eqn:trans}).
In Eq. (\ref{eqn:O2}), there are both creation and annihilation processes of the excitation modes.
Since we are interested in the creation process in Eq. (\ref{eqn:W}), we can express
\begin{align}
&W^x(\bq) = \frac{i}{2\sqrt{2}}
( u_\bq + v_{\bq} ) ( \alpha_{-\bq+}^\dagger - \alpha_{-\bq-}^\dagger ), \cr
&W^y(\bq) = \frac{1}{2\sqrt{2}}
( u_\bq + v_\bq ) ( \alpha_{-\bq+}^\dagger + \alpha_{-\bq-}^\dagger ), \label{eqn:O-q} \\
&W^z(\bq) =
-\frac{i}{2} ( u_\bq + v_\bq ) \alpha_{-\bq0}^\dagger.
\nonumber
\end{align}
Here, we used the relations $u_{-\bq}=u_\bq$ and $v_{-\bq}=v_\bq$.
Substituting Eq. (\ref{eqn:O-q}) into Eq. (\ref{eqn:P0}), we obtain
\begin{align}
P^\alpha
&= \frac{\sqrt{N}}{2\sqrt{2}}( u_0 + v_0 )
\Bigl\{
   \left[ iC^\alpha_x(+) + C^\alpha_y(+) \right] \alpha_{0+}^\dagger \cr
&~~~~~~~~~~~~~~~~~~~~~~~
+ \left[ -i\sqrt{2} C^\alpha_z(+) \right] \alpha_{00}^\dagger \cr
&~~~~~~~~~~~~~~~~~~~~~~~
+ \left[ -iC^\alpha_x(+) + C^\alpha_y(+) \right] \alpha_{0-}^\dagger \Bigr\} \cr
&+ \frac{\sqrt{N}}{2\sqrt{2}}( u_{\bk_\rB} + v_{\bk_\rB} )
\Bigl\{
  \left[ iC^\alpha_x(-) + C^\alpha_y(-) \right] \alpha_{\bk_\rB +}^\dagger \cr
&~~~~~~~~~~~~~~~~~~~~~~~
+ \left[ -i\sqrt{2} C^\alpha_z(-) \right] \alpha_{\bk_\rB 0}^\dagger \cr
&~~~~~~~~~~~~~~~~~~~~~~~
+ \left[ -iC^\alpha_x(-) + C^\alpha_y(-) \right] \alpha_{\bk_\rB -}^\dagger \Bigr\}. \cr
\label{eqn:p02}
\end{align}
The wavelength of the microwave is much larger than the sample size,
and the electric field of the microwave can be treated as a uniform one.
On the other hand, the coefficient tensor for the electric dipole has both the uniform and staggered components
with respect to the A and B sublattices.
The $C(+)$ term causes an excitation at $\bk=0$, whereas the $C(-)$ term is for $\bk=\bk_\rB$
\cite{Kimura-2018}.
Therefore, six excitation modes are observable [see Eq. (\ref{eqn:p02})].
Three of them are for $\bk=0$ and the ESR intensity is proportional to $(u_0+v_0)^2$.
The rest three are for $\bk_\rB=(0,0,2\pi)$ and the intensity is proportional to $(u_{\bk_\rB}+v_{\bk_\rB})^2$.
The values are expressed as
\begin{equation}
\begin{aligned}
&(u_0+v_0)^2 = \frac{\epsilon_0 + \gamma_0}{E_0}
= \sqrt{ \frac{J - (J_1 + J_2 + 2J_3)}{J} }
= \frac{E_{0}}{J}, \cr
&(u_{\bk_\rB}+v_{\bk_\rB})^2 = \frac{\epsilon_{\bk_\rB} + \gamma_{\bk_\rB}}{E_{\bk_\rB}}
= \sqrt{ \frac{J - (J_1 + J_2 - 2J_3)}{J} }
= \frac{E_{\bk_\rB}}{J}.
\end{aligned}
\label{eqn:factor}
\end{equation}
Notice that the intensities are independent on the magnetic field.
In Eq. (\ref{eqn:factor}), $E_0$ and $E_{\bk_\rB}$ are excitation energies
defined by Eq. (\ref{eqn:E-k}) for $\bk=0$ and $\bk=\bk_\rB$, respectively.
Notice that $E_0$ coincides with the excitation gap at $H=0$.
When the pressure is increased, $E_0$ decreases and becomes soft at the critical pressure.
Therefore, $(u_0+v_0)^2 = 0$ at the critical pressure
and the intensity for $\bk=0$ (A mode) vanishes at $H=0$.
In contrast to this, $E_{\bk_\rB}$ does not become zero
and the intensity for $\bk=\bk_\rB$ (B mode) is finite at the critical pressure.
For \K~and \Tl, we list the values of $(u_\bq+v_\bq)^2$ at the ambient pressure in Table \ref{table:uv}.
We can see that the value of $(u_0+v_0)^2$ is reduced in \Tl,
since it is close to the quantum critical point at the ambient pressure.
This explains why \K~possesses much stronger intensity of the direct transition than \Tl.

%%%%%%%%%%%%%%%%%%%%%%%%%%%%%%%%%%%%%%%%%%%%%%%%%%%%%%%%%%%%%%%%%%%%%%%%
\begin{table}[t]
\begin{center}
\caption{
Values of $(u_\bq+v_\bq)^2$ for \K~at the ambient pressure.
Exchange interaction parameters in Table \ref{table:J} are used.
Values for \Tl~are also shown here for comparison.
}
\label{table:uv}
\begin{tabular}{ccc}
\hline
Material & $\bq$ & $(u_\bq+v_\bq)^2=E_\bq/J$ \cr
\hline
\K  & $0$   & $0.610$ \cr
      & $\bk_\rB$ & $1.029$ \cr
\hline
\Tl & $0$   & $0.129$ \cr
      & $\bk_\rB$ & $0.824$ \cr
\hline    
\end{tabular}
\end{center}
\end{table}
%%%%%%%%%%%%%%%%%%%%%%%%%%%%%%%%%%%%%%%%%%%%%%%%%%%%%%%%%%%%%%%%%%%%%%%%

%%%%%%%%%%%%%%%%%%%%%%%%%%%%%%%%%%%%%%%%%%%%%%%%%%%%%%%%%%%%%%%%%%%%%%%%
\subsubsection{$\bH \perp (10\bar{2})$}
\label{sec:1}
%%%%%%%%%%%%%%%%%%%%%%%%%%%%%%%%%%%%%%%%%%%%%%%%%%%%%%%%%%%%%%%%%%%%%%%%

Let us first discuss the ESR intensity under the field applied along the crystal $z$-axis
(see Fig. \ref{fig:coordinate}) [$\bH\perp (10\bar{2})$].
In this case, the $z$-axis of the spin space is taken parallel to the field [see Eqs. (\ref{eqn:E-k}) and (\ref{eqn:p02})].
The coefficient tensor given by Eq. (\ref{eqn:C-AB}) is introduced in the same coordinates for the spin space.
Thus, the ESR intensity can be obtained by substituting Eq. (\ref{eqn:p02}) into Eq. (\ref{eqn:W})
with the coefficient tensor in Eq. (\ref{eqn:C-pm}).

%%%%%%%%%%%%%%%%%%%%%%%%%%%%%%%%%%%%%%%%%%%%%%%%%%%%%%%%%%%%%%%%%%%%%%%%
\subsubsection{$\bH \parallel [201]$}
%%%%%%%%%%%%%%%%%%%%%%%%%%%%%%%%%%%%%%%%%%%%%%%%%%%%%%%%%%%%%%%%%%%%%%%%

We next consider a case of the field applied in the [201] direction ($x$ direction).
The $z$-axis of the spin space is taken parallel to the field.
It is different from that of the crystal coordinate shown in Fig. \ref{fig:coordinate}.
For $\bH\parallel x$, the vector spin chirality in Eq. (\ref{eqn:P0}) is written as
\begin{align}
\begin{pmatrix}
W^x \cr
W^y \cr
W^z
\end{pmatrix}
&=
\begin{pmatrix}
\tilde{W}^z \cr
\tilde{W}^x \cr
\tilde{W}^y
\end{pmatrix} \cr
&=
\begin{pmatrix}
0 & 0 & 1 \cr
1 & 0 & 0 \cr
0 & 1 & 0
\end{pmatrix}
\begin{pmatrix}
\tilde{W}^x \cr
\tilde{W}^y \cr
\tilde{W}^z
\end{pmatrix} \cr
&\equiv R
\begin{pmatrix}
\tilde{W}^x \cr
\tilde{W}^y \cr
\tilde{W}^z
\end{pmatrix}.
\end{align}
Here, $\tilde{\bW}$ is the vector spin chirality in the transformed coordinates of the spin space,
where $\tilde{W}^z$ is taken parallel to the external magnetic field.
$R$ is a matrix for $2\pi/3$ rotation around the [111] direction.
The perturbation Hamiltonian in Eq. (\ref{eqn:Hp}) is then written as
\begin{align}
\H'_P
&=-\bE^\omega\cdot\bP \cr
&=-E^\omega_\alpha C^\alpha_\beta W^\beta \cr
&=-E^\omega_\alpha C^\alpha_\beta R_{\beta\gamma} \tilde{W}^\gamma \cr
&=-E^\omega_\alpha \tilde{C}^\alpha_\gamma \tilde{W}^\gamma.
\end{align}
Here, $\tilde{C}$ is the coefficient tensor for $\tilde{W}$.
For $\H\parallel [201]$, it is given by
\begin{equation}
\begin{aligned}
&\tilde{C}_{\bH \parallel [201]}(+) = C(+) R =
\begin{pmatrix}
0 & C^x_z &C^x_x \cr
C^y_y & 0 & 0 \cr
0 & C^z_z & C^z_x
\end{pmatrix}, \cr
&C_{\bH \parallel [201]}(-) = C(-) R =
\begin{pmatrix}
C^x_y & 0 & 0 \cr
0 & C^y_z & C^y_x \cr
C^z_y & 0 & 0
\end{pmatrix},
\end{aligned}
\label{eqn:C-pm-new1}
\end{equation}
The matrix element for the transition can be obtained by Eq. (\ref{eqn:p02})
with the transformed coefficient tensor in Eq. (\ref{eqn:C-pm-new1}) instead of Eq. (\ref{eqn:C-pm}).

%%%%%%%%%%%%%%%%%%%%%%%%%%%%%%%%%%%%%%%%%%%%%%%%%%%%%%%%%%%%%%%%%%%%%%%%
\subsubsection{$\bH \parallel [010]$}
%%%%%%%%%%%%%%%%%%%%%%%%%%%%%%%%%%%%%%%%%%%%%%%%%%%%%%%%%%%%%%%%%%%%%%%%

For $\bH \parallel [010]$, the effective coefficient tensor is obtained in the same manner as in the $\bH\parallel [201]$ case.
For $\bH \parallel [010]$, it is given by
\begin{equation}
\begin{aligned}
&\tilde{C}_{\bH \parallel [010]}(+) = C(+) R^{-1} =
\begin{pmatrix}
C^x_z & C^x_x & 0 \cr
0 & 0 & C^y_y \cr
C^z_z & C^z_x & 0
\end{pmatrix}, \cr
&\tilde{C}_{\bH \parallel [010]}(-) = C(-) R^{-1} =
\begin{pmatrix}
0 & 0 & C^x_y \cr
C^y_z & C^y_x & 0 \cr
0 & 0 & C^z_y
\end{pmatrix}.
\end{aligned}
\label{eqn:C-pm-new2}
\end{equation}
The matrix element for the transition can be obtained by Eq. (\ref{eqn:p02})
with the transformed coefficient tensor in Eq. (\ref{eqn:C-pm-new2}) instead of Eq. (\ref{eqn:C-pm}).

In Table \ref{table:W-linear}, we summarize the matrix elements for ESR in the disordered phase
under the various directions of the external magnetic field.

%%%%%%%%%%%%%%%%%%%%%%%%%%%%%%%%%%%%%%%%%%%%%%%%%%%%%%%%%%%%%%%%%%%%%%%%
\begin{table*}[t]
\begin{center}
\caption{
Matrix elements for ESR with linearly polarize light under various directions of the external magnetic field.
The matrix element is defined by $\braket{{\rm ES}|\H'_P|{\rm GS}}$.
Here, $|{\rm GS}\rangle$ and $|{\rm ES}\rangle$ represent the ground and excited states, respectively.
In $\H'_P$, we use the expression of the total electric dipole given by in Eq. (\ref{eqn:p02}).
The coefficient tensors are shown in Eqs. (\ref{eqn:C-pm}), (\ref{eqn:C-pm-new1}), and (\ref{eqn:C-pm-new2})
for $\bH\perp (10\bar{2})$, $\bH\parallel [201]$, and $\bH\parallel [010]$, respectively.
The energy for the excitation mode $\alpha_{\bq m}$ decreases (increases) with the field for $m=+$ ($m=-$),
while it is constant for $m=0$.
The $x$, $y$, and $z$ directions for the electric field of the microwave and the coefficient tensors
are defined in the crystal coordinates shown in Fig. \ref{fig:coordinate}.
As in Eq. (\ref{eqn:p02}), the prefactor of the matrix element is
$\frac{\sqrt{N}}{2\sqrt{2}} \sqrt{\frac{E_0}{J}}$ and $\frac{\sqrt{N}}{2\sqrt{2}} \sqrt{\frac{E_{\bk_\rB}}{J}}$
for the excited states at $\bq=0$ (A mode) and $\bq=\bk_\rB$ (B mode),
respectively [see Eqs. (\ref{eqn:p02}) and (\ref{eqn:factor})].
For names of the excitation modes, see Figs. \ref{fig:FH}(a) and \ref{fig:FH}(b).
At the ambient pressure, the values of $E_\bq/J$ are shown in Table \ref{table:uv}.
ESR intensity is obtained by taking square of the absolute value of the matrix element.
}
\label{table:W-linear}
\begin{tabular}{clc}
\hline
Field direction & Excitation mode ($\alpha_{\bq m}$)  & Matrix element \cr
\hline
$\bH \perp (10\bar{2})$ & A$_-$ ($\alpha_{0+}$) & $i E^\omega_x C^x_x + E^\omega_y C^y_y + i E^\omega_z C^z_x$ \cr
                                             & A$_+$ ($\alpha_{0-}$)  & $- i E^\omega_x C^x_x + E^\omega_y C^y_y - i E^\omega_z C^z_x$ \cr
                                             & A$_0$ ($\alpha_{00}$) & $-i\sqrt{2} ( E^\omega_x C^x_z + E^\omega_z C^z_z )$ \cr
                                             & B$_-$ ($\alpha_{\bk_\rB +}$) & $E^\omega_x C^x_y + i E^\omega_y C^y_x + E^\omega_z C^z_y$ \cr
                                             & B$_+$ ($\alpha_{\bk_\rB -}$) &  $E^\omega_x C^x_y - i E^\omega_y C^y_x + E^\omega_z C^z_y$ \cr
                                             & B$_0$ ($\alpha_{\bk_\rB 0}$) &  $-i\sqrt{2} E^\omega_y C^y_z$ \cr
\hline
$\bH \parallel [201]$ & A$_-$ ($\alpha_{0+}$) & $E^\omega_x C^x_z + i E^\omega_y C^y_y + E^\omega_z C^z_z$ \cr
                                       & A$_+$ ($\alpha_{0-}$) &  $E^\omega_x C^x_z - i E^\omega_y C^y_y + E^\omega_z C^z_z$ \cr
                                       & A$_0$ ($\alpha_{00}$) & $-i\sqrt{2} ( E^\omega_x C^x_x + E^\omega_z C^z_x )$ \cr
                                       & B$_-$ ($\alpha_{\bk_\rB +}$) & $i E^\omega_x C^x_y + E^\omega_y C^y_z + i E^\omega_z C^z_y$ \cr
                                       & B$_+$ ($\alpha_{\bk_\rB -}$)  & $-i E^\omega_x C^x_y + E^\omega_y C^y_z - i E^\omega_z C^z_y$ \cr
                                       & B$_0$ ($\alpha_{\bk_\rB 0}$) & $-i\sqrt{2} E^\omega_y C^y_x$ \cr
\hline
$\bH \parallel [010]$ & A$_-$ ($\alpha_{0+}$) & $E^\omega_x ( i C^x_z + C^x_x ) + E^\omega_z ( i C^z_z + C^z_x )$ \cr
                                       & A$_+$ ($\alpha_{0-}$)  & $E^\omega_x ( - i C^x_z + C^x_x ) + E^\omega_z ( - i C^z_z + C^z_x )$ \cr
                                       & A$_0$ ($\alpha_{00}$) & $-i\sqrt{2} E^\omega_y C^y_y$ \cr
                                       & B$_-$ ($\alpha_{\bk_\rB +}$) & $E^\omega_y ( i C^y_z + C^y_x )$ \cr
                                       & B$_+$ ($\alpha_{\bk_\rB -}$)  & $E^\omega_y ( - i C^y_z + C^y_x )$ \cr
                                       & B$_0$ ($\alpha_{\bk_\rB 0}$) & $-i\sqrt{2} ( E^\omega_x C^x_y + E^\omega_z C^z_y )$ \cr
\hline
\end{tabular}
\end{center}
\end{table*}
%%%%%%%%%%%%%%%%%%%%%%%%%%%%%%%%%%%%%%%%%%%%%%%%%%%%%%%%%%%%%%%%%%%%%%%%

%%%%%%%%%%%%%%%%%%%%%%%%%%%%%%%%%%%%%%%%%%%%%%%%%%%%%%%%%%%%%%%%%%%%%%%%
\section{Experimental Results}
%%%%%%%%%%%%%%%%%%%%%%%%%%%%%%%%%%%%%%%%%%%%%%%%%%%%%%%%%%%%%%%%%%%%%%%%

High-pressure THz ESR measurements have been performed by our recently developed setup
\cite{Sakurai-2015}.
The hydrostatic pressure is generated by the specially designed pressure cell for THz ESR measurement.
Two single crystals with the dimensions of 2$\times$1$\times$5 mm$^{3}$ cut along the [201] directions were used.
The external magnetic field was applied parallel to the [201] direction and the temperature is 2 K.
The maximum field is 10 T and the frequency region is from 0.1 to 0.7 THz.
Daphne 7373 is used as the pressure-transmitting fluid and the maximum pressure is 1 GPa in this study.
The pressure is estimated by the relation between the load at room temperature and the pressure at low temperatures
\cite{Sakurai-2015}.

%%%%%%%%%%%%%%%%%%%%%%%%%%%%%%%%%%%%%%%%%%%%%%%%%%%%%%%%%%%%%%%%%%%%%%%%
\subsection{Below the Critical Pressure ($P<\Pc$)}
%%%%%%%%%%%%%%%%%%%%%%%%%%%%%%%%%%%%%%%%%%%%%%%%%%%%%%%%%%%%%%%%%%%%%%%%

%%%%%%%%%%%%%%%%%%%%%%%%%%%%%%%%%%%%%%%%%%%%%%%%%%%%%%%%%%%%%%%%%%%%%%%%
\begin{figure*}
\begin{center}
\includegraphics[width=12cm]{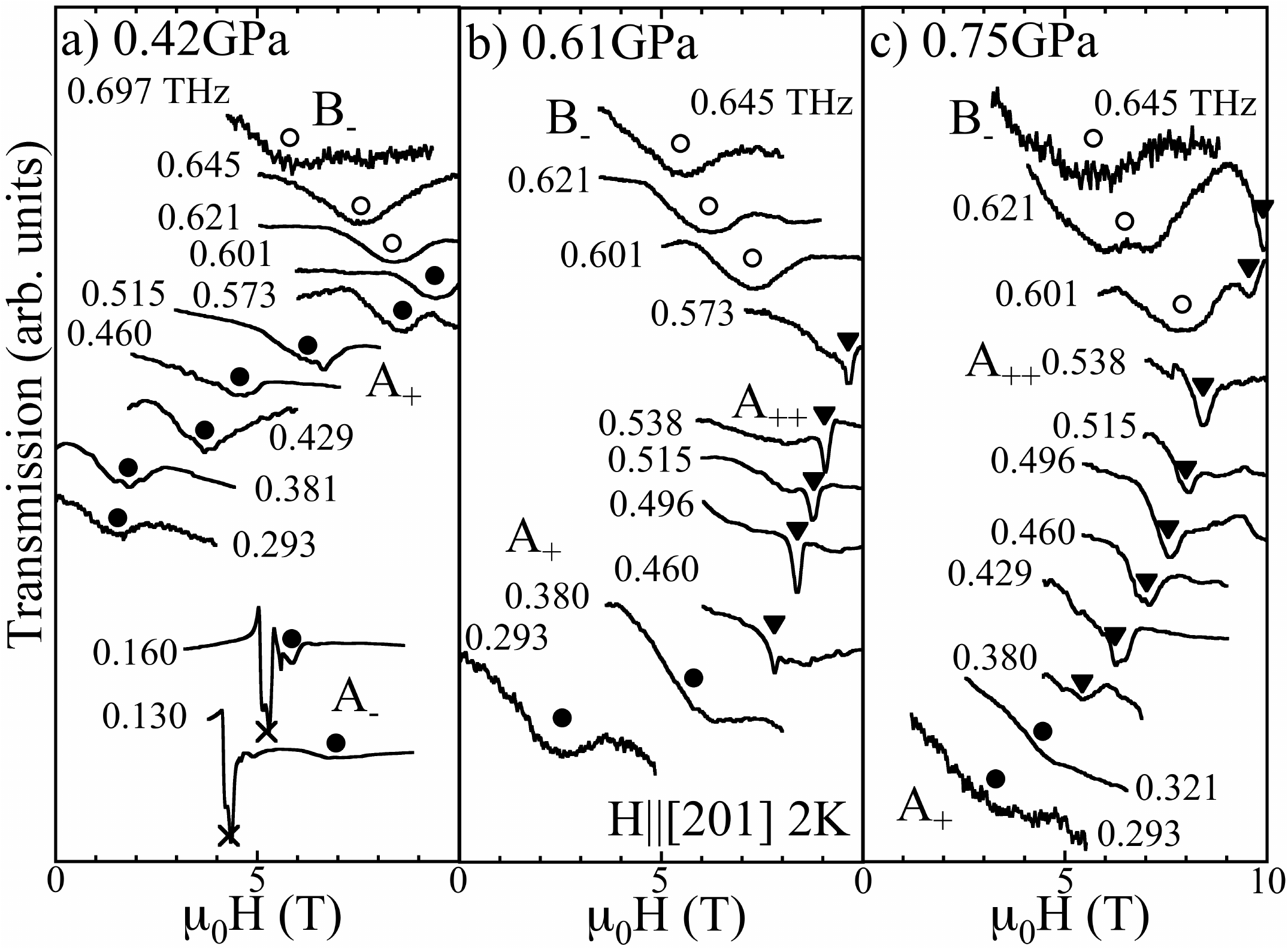}
\caption{
Frequency dependence of the ESR spectra at 0.42 GPa (a), 0.61 GPa (b), and 0.75 GPa (c),
where the pressures are below the critical pressure $(P<\Pc$).
The A and B modes were observed (see text for details).
At 0.42 GPa, the absorption lines indicated by cross come from impurities.
}
\label{fig:nama-data}
\end{center}
\end{figure*}
%%%%%%%%%%%%%%%%%%%%%%%%%%%%%%%%%%%%%%%%%%%%%%%%%%%%%%%%%%%%%%%%%%%%%%%%

%%%%%%%%%%%%%%%%%%%%%%%%%%%%%%%%%%%%%%%%%%%%%%%%%%%%%%%%%%%%%%%%%%%%%%%%
\begin{figure*}
\begin{center}
\includegraphics[width=12cm]{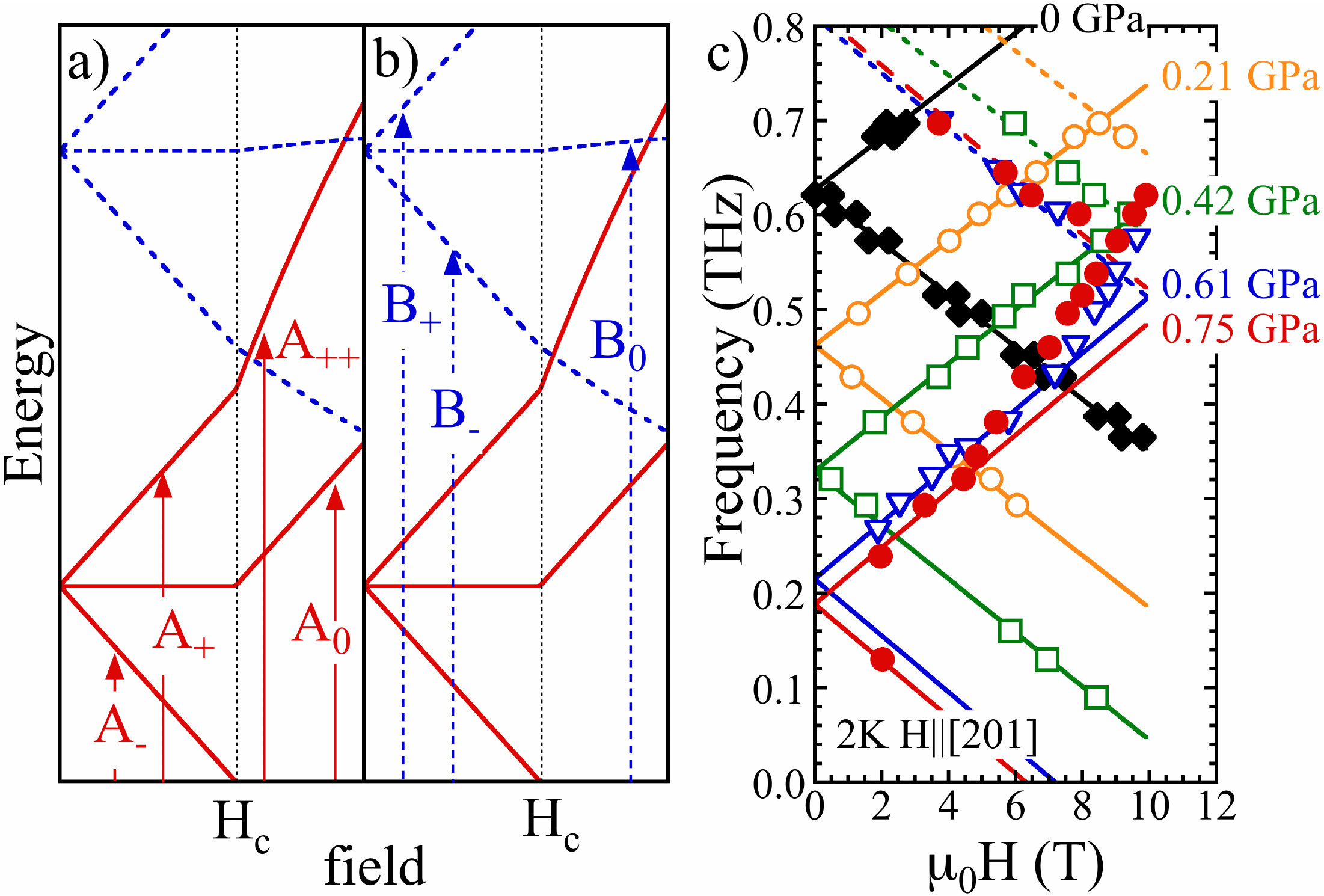}
\caption{
(Color online)
Schematic frequency-field diagram observed at the ambient pressure for (a) A modes and (b) B modes
\cite{Tanaka-1998,Kimura-2004}.
Notice that the two modes do not directly correspond to the A and B dimer sublattices shown in Fig. \ref{fig:interaction}, respectively,
but are originating from the uniform (A$+$B) and staggered (A$-$B) components of the electric dipole,
respectively (see text for details).
The subscript $\pm$ indicate the transitions from the ground state to the $S_{z}=\mp 1$ triplet states, respectively.
In the same way, A$_0$ and B$_0$ modes are for $S_{z}=0$.
(c) Experimentally obtained spectra at various pressures in the present study.
The $g$ factor for the A$_\pm$ modes is $g\simeq 2.05$,
whereas it is $g_{\rm eff}\simeq 4.22$ for the A$_{++}$ mode at 0.61 GPa.
}
\label{fig:FH}
\end{center}
\end{figure*}
%%%%%%%%%%%%%%%%%%%%%%%%%%%%%%%%%%%%%%%%%%%%%%%%%%%%%%%%%%%%%%%%%%%%%%%%

Figure \ref{fig:nama-data} shows the typical ESR spectra obtained under pressures.
We observed two gapped ESR modes denoted by A (filled symbols) and B (open circle).
It is known that the A and B modes are the transitions from the singlet ground state to the triplet states ($S=1$),
and the energy-field diagram is schematically shown in Figs. \ref{fig:FH} (a) and \ref{fig:FH} (b)
\cite{Tanaka-1998,Kimura-2004}.

In the disordered phase, the transitions from the singlet to the $S_z=\pm 1$ triplet states are forbidden
in the conventional interpretation of the ESR spectrum by the magnetic dipole transition.
Nevertheless, such excitation modes have been clearly observed.
This problem was investigated from a viewpoint of a magnetoelectric effect,
and it was recently revealed that the transitions are caused by the electric-field component of the microwave
through an electric dipole generated by the vector spin chirality $\bS_l \times \bS_r$ of a dimer
\cite{Kimura-2018}.
It was also reported that the vector spin chirality induces ferroelectricity
in the field-induced ordered phase in \Tl~\cite{Kimura-2016,Kimura-2017}.
Owing to the A and B dimer sites at the corner and the center in a unit cell (see Fig. \ref{fig:interaction}),
the electric dipole has both uniform and staggered components with respect to the A and B dimer sites.
These components enable the detection of the A and B modes in the ESR measurements
\cite{Kimura-2018}.
The excitation energy of the A mode at $H=0$ corresponds to the energy
at the band minimum $\bk_{\rm A} = (0,0,0)(=\bQ)$ in the magnon band \cite{Cavadini-1999,note:Q},
whereas those of the B mode corresponds to that at $\bk_{\rm B}=(0,0,2\pi)$.
It was confirmed that the former and the latter modes
are due to the uniform and staggered components of the electric dipole, respectively
\cite{Kimura-2018}.

In Fig. \ref{fig:nama-data}, it is clear that there is an apparent pressure effect in the spectrum.
Comparing the observed resonance fields in Fig. \ref{fig:nama-data} and the diagram in Fig. \ref{fig:FH}(a),
we assign the A modes and plot them in Fig. \ref{fig:FH}(c) for various pressures.
Below the critical field, the A$_{\pm}$ modes have a linear field dependence.
The $g$-factor of the modes agree within $g=2.05\pm 0.03$.
In Fig. \ref{fig:FH}(c), it is clear that the excitation energy at $H=0$ is reduced by applying the pressure.
As is seen in the ESR spectra at 0.61 GPa [Fig. \ref{fig:nama-data} (b)],
the A$_+$ mode drastically changes to the A$_{++}$ mode at 0.496 THz (7.9 T).
This mode corresponds to the A$_{++}$ mode shown in Fig. \ref{fig:FH}(a).
The A$_{++}$ mode in Fig. \ref{fig:nama-data} (b) shows very narrow line above 0.496 THz
compared with that of the A$_{+}$ mode.
This sudden change at 0.496 THz (7.9 T) is due to the field-induced phase transition at the critical field $H_{\rm c}$. 
The slope of the A$_{++}$ mode is almost twice ($g_{\rm eff}=4.22$) as that of the A$_+$ mode below $H_{\rm c}$,
as shown in Fig. \ref{fig:FH}(c) for 0.61 GPa.
This is because the $S_{z} = 1$ state merges into the singlet ground state above $H_{\rm c}$
\cite{Matsumoto-2004}.

\begin{figure}
\begin{center}
\includegraphics[width=8cm]{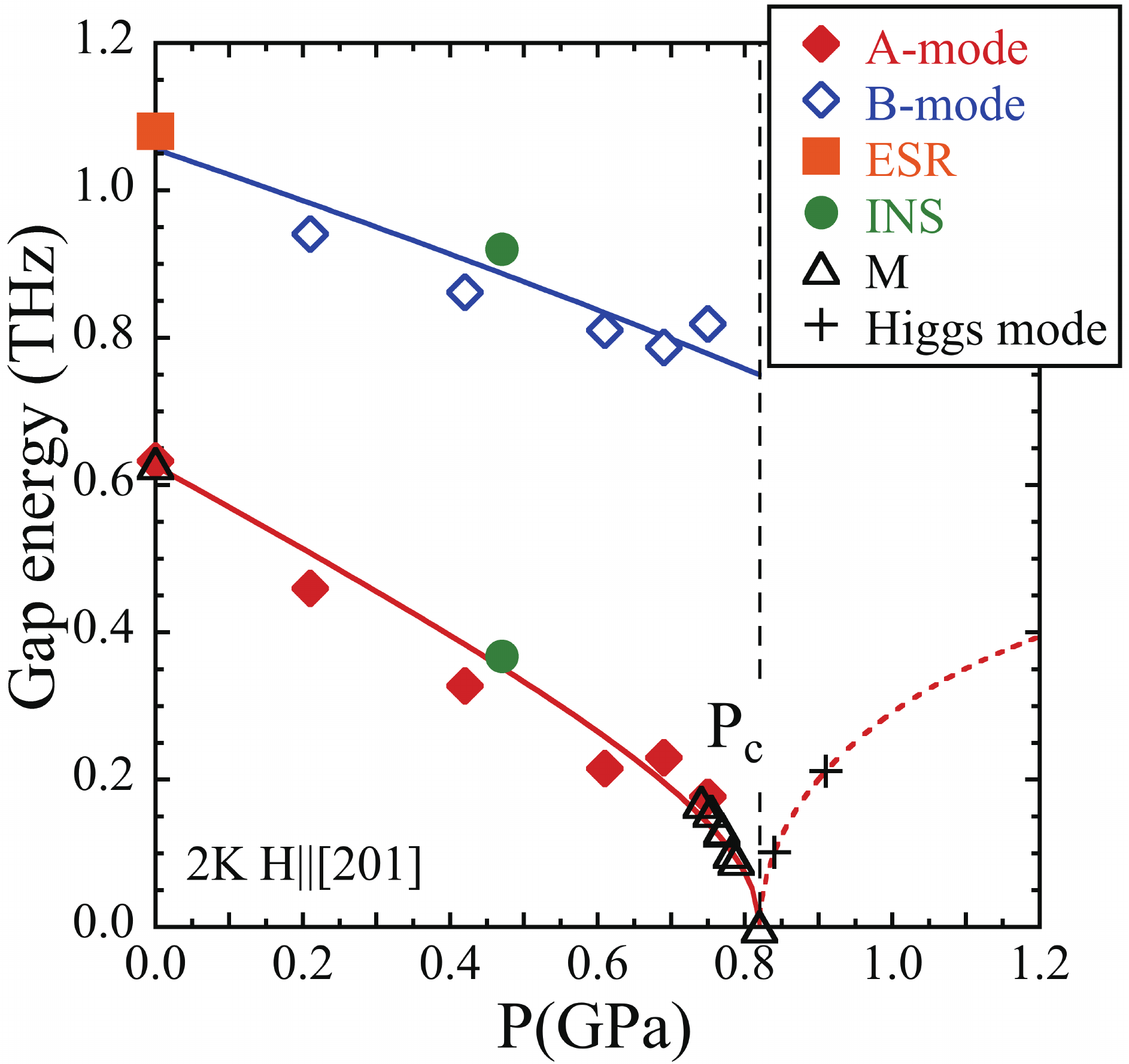}
\caption{
(Color online)
Pressure dependence of the excitation energies.
The symbols A-mode and B-mode represent the data obtained in the present study.
The temperature is 2 K and the external magnetic field is applied in the [201] direction.
The symbol ESR represents the data from the previous ESR measurement at the ambient pressure
\cite{Kimura-2004}.
The symbol INS and M represent the data from the inelastic neutron scattering
\cite{Goto-2007}
and the magnetization
\cite{Goto-2006}
measurements under pressures, respectively.
The solid curves are the fitting results with Eqs. (\ref{eqn:EA}) and (\ref{eqn:EB}).
The dashed line is the calculated pressure dependence of the energy gap of the Higgs amplitude mode [Eq. (\ref{eqn:Higgs})].
For the pressure dependences of the exchange interactions, Eq. (\ref{eqn:parameter}) is used
with the parameters shown in Tables \ref{table:J} and \ref{table:parameter}.
Above $\Pc$, the observed energy gap of the Higgs amplitude mode is shown with symbol ``+",
which are denoted by $E_{\rm L}(P)$ in Figs. \ref{fig:higgs}(c) and \ref{fig:higgs}(d).
}
\label{fig:gapP}
\end{center}
\end{figure}

The excitation energy of the A mode at $H=0$ can be estimated by the linear extrapolation at each pressure.
In the same way, that of the B mode is also estimated
on the assumption that the $g$-factor is the same as that of the A mode.
The excitation energies of the A and B modes at $H=0$ are plotted against the pressure in Fig. \ref{fig:gapP}
together with the data taken from the magnetization
\cite{Goto-2006}
and the inelastic neutron scattering
\cite{Goto-2007}
measurements under pressure.
In Fig. \ref{fig:gapP}, we can see how the excitation energies of the A and B modes evolve with the pressure.
Both energies decrease with the pressure, where the former decreases more quickly than the latter.

%%%%%%%%%%%%%%%%%%%%%%%%%%%%%%%%%%%%%%%%%%%%%%%%%%%%%%%%%%%%%%%%%%%%%%%%
%%%%%%%%%%%%%%%%%%%%%%%%%%%%%%%%%%%%%%%%%%%%%%%%%%%%%%%%%%%%%%%%%%%%%%%%

Next, we extract pressure dependences of the intradimer and interdimer interactions from our observed data.
The magnon dispersion relation at the ambient pressure is given by Eq. (\ref{eqn:E-k}).
At pressure $P$, it is replaced as
\begin{align}
E_\bk(P) =\sqrt{ J^2(P) + 2 J(P) \gamma_\bk(P) },
\label{eqn:E-P}
\end{align}
in the absence of the magnetic field.
Here, $J(P)$ is the intradimer interaction at pressure $P$.
$\gamma_\bk(P)$ is expressed by Eq. (\ref{eqn:gamma}) with $J_{i}(P)$ ($i = 1, 2, 3$)
as the interdimer interactions at pressure $P$.
The values of $\gamma_\bk(P)$ at $\boldsymbol{k}_{\rm A}=(0, 0, 0)$ and $\boldsymbol{k}_{\rm B}=(0, 0, 2\pi)$ are given by
\begin{align}
\gamma_{\bk_{\rm A}}(P) &= -\frac{1}{2} \left[ J_1(P) + J_2(P) + 2J_3(P) \right], \\
\gamma_{\bk_{\rm B}}(P) &= -\frac{1}{2} \left[ J_1(P) + J_2(P) - 2J_3(P) \right].
\end{align}
We can see that the sign of $J_3(P)$ is different in these equations.
This explains the different pressure dependences of the excitation energies of the A and B modes shown in Fig. \ref{fig:gapP}.

For the pressure dependences of the exchange interactions,
we assume that $J(P)$, $J_1(P) + J_2(P)$, and $J_3(P)$ have the following linear $P$ dependences:
\begin{align}
J(P) &= s P + J(0), \cr
J_1(P) + J_2(P) &= t P + J_1(0) + J_2(0),
\label{eqn:parameter} \\
J_3(P) &= u P + J_3(0).
\nonumber
\end{align}
Here, $s$, $t$, and $u$ are linear coefficients for the $P$ dependence.
The excitation energies of the A and B modes are then expressed as
\begin{align}
E_{\bk_{\rm A}}(P)
&= \sqrt{ \left[ J(0) + s P \right] \left\{ \left[ J(0) + 2\gamma_{\bk_{\rm A}}(0) \right] + \left( s - t - 2u \right) P \right\} },
\label{eqn:EA} \\
E_{\bk_{\rm B}}(P)
&= \sqrt{ \left[ J(0) + s P \right] \left\{ \left[ J(0) + 2\gamma_{\bk_{\rm B}}(0) \right] + \left( s - t + 2u \right) P \right\} }.
\label{eqn:EB}
\end{align}
By fitting the observed results shown in Fig. \ref{fig:gapP} with Eqs. (\ref{eqn:EA}) and (\ref{eqn:EB}),
we can extract the coefficient parameters $s$, $t$, and $u$.
First, the exchange interactions at $P=0$ GPa are already known
and are listed in Table. \ref{table:J}.
Next, we determine the parameters $s$ and $q=s-t-2u$
by fitting Eq. (\ref{eqn:EA}) to the pressure dependence of the excitation energy of the A mode, as shown in Fig. \ref{fig:gapP}.
The parameters are determined to be $s=-0.562$ THz/GPa and $q=-0.465$ THz/GPa.
The critical pressure $P_{c}$ is also determined to be $P_{c}=0.82$ GPa,
which is consistent with the value obtained by the magnetization measurement under pressure
\cite{Goto-2006}.
Then, using the obtained parameter $s$, we determine the parameter $r=s-t+2u$
by fitting Eq. (\ref{eqn:EB}) to the pressure dependence of the excitation energy of the B mode.
The result is $r=-0.113$ THz/GPa.
By solving the equations $t+2u=s-q$ and $t-2u=s-r$, the parameters $t$ and $u$ are obtained.
Thus, we succeeded in separating the pressure dependences of the exchange interactions.

We summarize the obtained linear coefficients in Table \ref{table:parameter}.
We can see that $J$ and $J_1+J_2$ decrease with pressure, whereas $J_3$ slightly increases.
The intradimer interaction $J$ has the strongest pressure dependence,
and this is the main origin of the reduction of the excitation energies.
On the other hand, $J_1+J_2$ decreases with pressure and this suppresses the reduction of the excitation energies.
Moreover, it is found that $J_3$ contributes to the reduction of the excitation energy of the A mode,
while it suppresses the reduction for the B mode. 

%%%%%%%%%%%%%%%%%%%%%%%%%%%%%%%%%%%%%%%%%%%%%%%%%%%%%%%%%%%%%%%%%%%%%%%%
\begin{table}
\caption{
Parameters for the linear $P$ dependence of the exchange interactions.
From the pressure dependence of the excitation gap, we can only extract the pressure dependence of $J_1+J_2$.
Taking into account the magnetic field dependence of the excitation modes under various pressures,
we extracted the following pressure dependences of $J_1$ and $J_2$.
Here, $J_1(P)=t_1 P+J_1(0)$ and $J_2(P)=t_2 P+J_2(0)$.
In the extended spin-wave theory, we use the obtained linear $P$ dependence of the interactions
together with the isotropic $g$-factor $g=2.05$.
}
\label{table:parameter}
\begin{center}
\begin{tabular}{ccc}
\hline
Exchange interaction & Parameter & Value (THz/GPa) \\
\hline
$J$               & $s$      & $-0.562$ \\
$J_1+J_2$ & $t$      & $-0.273$ \\
$J_1$          & $t_1$ & $-0.070$ \\
$J_2$          & $t_2$ & $-0.203$ \\
$J_3$          & $u$     & $+0.088$ \\
\hline
\end{tabular}
\end{center}
\end{table}
%%%%%%%%%%%%%%%%%%%%%%%%%%%%%%%%%%%%%%%%%%%%%%%%%%%%%%%%%%%%%%%%%%%%%%%%

%%%%%%%%%%%%%%%%%%%%%%%%%%%%%%%%%%%%%%%%%%%%%%%%%%%%%%%%%%%%%%%%%%%%%%%%
\subsection{Above the Critical Pressure ($P>\Pc$)}
%%%%%%%%%%%%%%%%%%%%%%%%%%%%%%%%%%%%%%%%%%%%%%%%%%%%%%%%%%%%%%%%%%%%%%%%

Performing the ESR measurements under various pressures for $P<\Pc$,
we now know how the exchange interactions depend on the pressure.
Equation (\ref{eqn:parameter}) and the parameters shown in Tables \ref{table:J} and \ref{table:parameter}
provide further information on the pressure dependence and enable us to analyze the ESR spectra observed above $\Pc$.
As shown in Fig. \ref{fig:gapP}, the A mode becomes soft at $\Pc$.
Above $\Pc$, a finite antiferromagnetic moment appears and the threefold degeneracy is lifted.
One mode becomes the Higgs amplitude mode with acquiring a finite energy gap.
The energy gap (minimum excitation energy) of the Higgs amplitude is located at $\bk_{\rm A}$
and its pressure dependence is analytically expressed as
\cite{Matsumoto-2004}
\begin{align}
E_{\bk_{\rm A}}(P) = \sqrt{\gamma_{\bk_{\rm A}}^2(P)-J^2(P)} \left[ \equiv E_{\rm L}(P) \right].
\label{eqn:Higgs}
\end{align}
This is plotted with the dashed line in Fig. \ref{fig:gapP} above $\Pc$,
as the continuation of the A mode after the softening.

\begin{figure}
\begin{center}
\includegraphics[width=8cm]{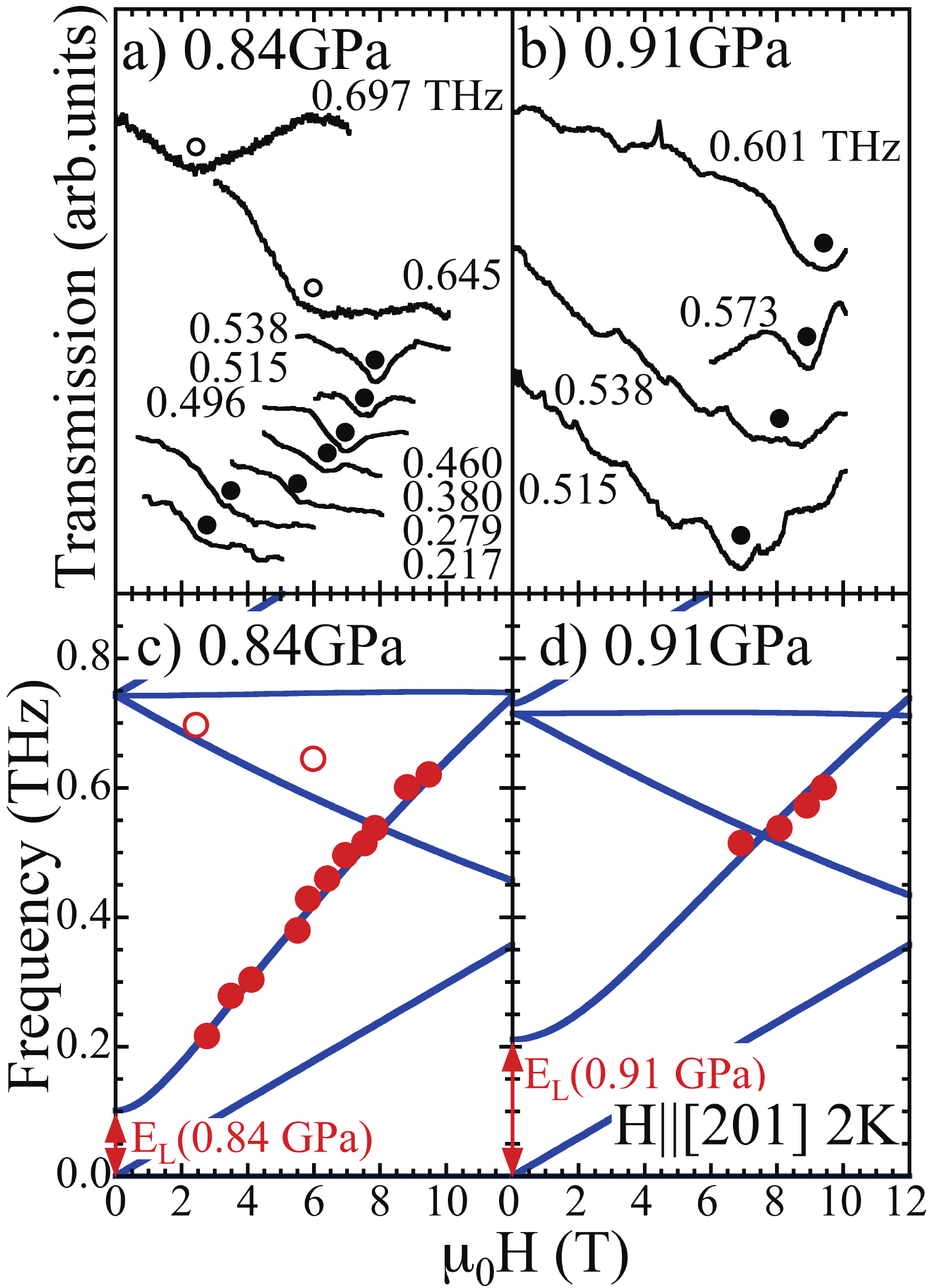}
\caption{
(Color online)
ESR spectra and the frequency-field diagram obtained above $P_{c}$.
(a) and (b) are ESR spectra obtained at 0.84 GPa and 0.91 GPa, respectively.
(c) and (d) are the corresponding frequency-field diagram.
$E_{\rm L}(P)$ indicates the energy gap of the longitudinal (Higgs amplitude) mode for $H=0$.
Circles are the observed data, whereas lines are calculated result by the extended spin-wave theory.
Here, we used the pressure-dependent interaction parameters shown in Tables \ref{table:J} and \ref{table:parameter}.
To fit the experimental results, the $g$-factor is chosen as $g=2.13$ in Eqs. (\ref{eqn:H-intra}) and (\ref{eqn:g-isotropic}),
which is somewhat enhance from the value at the ambient pressure ($g=2.05$).
The magnetic field gradient is $g_{\rm eff}\simeq 4.36$ for 0.84 GPa,
which is comparable to the value observed in the field-induced ordered phase
$(g_{\rm eff}\simeq 4.22)$ shown in Fig. \ref{fig:FH}(c) for 0.61 GPa.
}
\label{fig:higgs}
\end{center}
\end{figure}

Now we know how the energy gap of  the Higgs amplitude mode evolves with pressure in \K.
To detect the Higgs amplitude mode, we performed ESR measurements
in the pressure-induced ordered phase ($P>\Pc$).
Figures \ref{fig:higgs} (a) and \ref{fig:higgs}(b) show the ESR spectra observed above $P_{c}(=0.82$ GPa).
The main absorption line (filled circle) at 0.84 GPa, which is just above $P_{c}$,
seem to have almost the same slope ($g_{\rm eff}=4.36$) of that of the A$_{++}$ mode ($g_{\rm eff}\simeq 4.22$)
which typically appears in the field-induced ordered phase
[see the A$_{++}$ mode at 0.61 GPa in Fig. \ref{fig:FH} (c)].
This indicates that the spectrum in Fig. \ref{fig:higgs}(a) is from the ordered phase.
Differing from the A$_{++}$ mode shown in Fig. \ref{fig:nama-data} (b),
the lines are broadened and the intensities seem to be weakened at 0.84 GPa.
Moreover, when the pressure is increased to 0.91 GPa,
the excitation gap develops, the linewidth becomes further broadened,
and the absorption line is hardly resolved at lower frequencies [see Fig. \ref{fig:higgs}(b)].
Thus, the characters of the spectra in Figs. \ref{fig:higgs}(a) and \ref{fig:higgs}(b) are qualitatively different
from those of the A$_{++}$ mode observed below $\Pc$.
All these findings coincide with features of the Higgs amplitude mode
observed by inelastic neutron scattering and Raman scattering measurements in \Tl
\cite{Ruegg-2008,Kuroe-2012},
and imply that the Higgs amplitude is detected by ESR in \K.

%%%%%%%%%%%%%%%%%%%%%%%%%%%%%%%%%%%%%%%%%%%%%%%%%%%%%%%%%%%%%%%%%%%%%%%%
%%%%%%%%%%%%%%%%%%%%%%%%%%%%%%%%%%%%%%%%%%%%%%%%%%%%%%%%%%%%%%%%%%%%%%%%

To check this, we calculate the field dependence of the excitation energies by the extended spin-wave theory.
For 0.84 GPa and 0.91 GPa, the energy gaps of the Higgs amplitude mode were indicated by the "+" symbol in Fig. \ref{fig:gapP}.
They are denoted by $E_{\rm L}(P)$ in Figs. \ref{fig:higgs} (c) and \ref{fig:higgs}(d).
The field evolution of the energy gap starts with these values.
We next plot the calculated field dependence of the energy gaps
together with the observed data in Figs. \ref{fig:higgs} (c) and \ref{fig:higgs}(d).
We can see that the theoretical curves are consistent with the observed data.
This implies that the Higgs amplitude mode is detected by ESR.

%%%%%%%%%%%%%%%%%%%%%%%%%%%%%%%%%%%%%%%%%%%%%%%%%%%%%%%%%%%%%%%%%%%%%%%%
\section{Analysis of ESR Intensity}
%%%%%%%%%%%%%%%%%%%%%%%%%%%%%%%%%%%%%%%%%%%%%%%%%%%%%%%%%%%%%%%%%%%%%%%%

In the disordered phase, the singlet--triplet transition was observed by ESR.
This indicates that the transition is not magnetic but it has an electric origin.
Therefore, the vector spin chirality plays an important role in the ESR in \K,
as pointed out in \Tl~\cite{Kimura-2018,Kimura-2020}.
In the present measurement for $P<\Pc$, we can see in Fig. \ref{fig:nama-data}
that the spectrum of the A$_{++}$ mode at 0.61 GPa clearly becomes sharp compared to that of the A$_+$ mode.
Here, the A$_{++}$ mode is in the ordered phase and is connected to the A$_+$ mode in the disordered phase.
On the other hand, such an effect was not clearly seen at the ambient pressure
\cite{Kimura-2004}.
Besides, we need evidence that the observed mode in Figs. \ref{fig:higgs}(c) and \ref{fig:higgs}(d)
is the one certainly connected to the Higgs amplitude mode at $H=0$ with a finite excitation gap.
To examine these points, we calculate the field dependence of the ESR intensity
under various pressures by the extended spin-wave theory.
For this calculation, we use Eqs. (\ref{eqn:Wee}), (\ref{eqn:s-mat}), and (\ref{eqn:p-mat}) in the Appendix.
In the calculation, the magnetic field is applied in the $x$-axis ([201] direction), as in the experimental setting.
In the ordered phase we assume that a staggered moment appears in the $y$ direction,
since it is reported in the isostructural compound \Tl~that the easy axis is close to the [201] direction
\cite{Tanaka-2001,Oosawa-2003-PRB}
and the second easy axis is parallel to the [010] direction ($y$-axis)
\cite{Yamada-2008}.
Under a magnetic field, the staggered moment tends to align perpendicular to the field.
For $\bH \parallel [201]$, which is close to the easy-axis,
the staggered moment is expected to be aligned in the second easy axis as in \Tl.

%%%%%%%%%%%%%%%%%%%%%%%%%%%%%%%%%%%%%%%%%%%%%%%%%%%%%%%%%%%%%%%%%%%%%%%%
\subsection{Magnetic Channel}
%%%%%%%%%%%%%%%%%%%%%%%%%%%%%%%%%%%%%%%%%%%%%%%%%%%%%%%%%%%%%%%%%%%%%%%%

First, we study the magnetic channel.
The matrix elements of the spin operators are shown in Eq. (\ref{eqn:S+}).
For $\bH\parallel x$, the role of the spin operators in Eq. (\ref{eqn:S+}) should be replaced as
$(S^x,S^y,S^z) \rightarrow (S^y,S^z,S^x)$, respectively.
In the ordered phase, the ground state and the A$_\pm$ excitation modes are described
by a superposition of the singlet $(|00\rangle)$ and $|1\pm1\rangle$ triplet states,
while the A$_0$ mode is pure $|10\rangle$ triplet state.
Here, the $S^z$ values of the triplet states are taken parallel to the external magnetic field.

\begin{figure}
\begin{center}
\includegraphics[width=5.9cm]{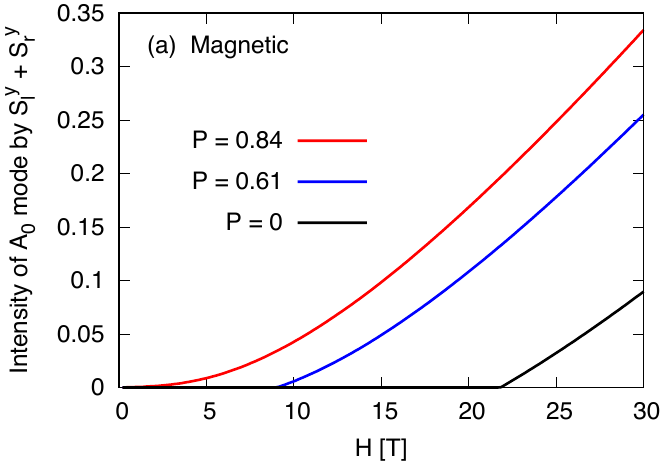}
\includegraphics[width=6cm]{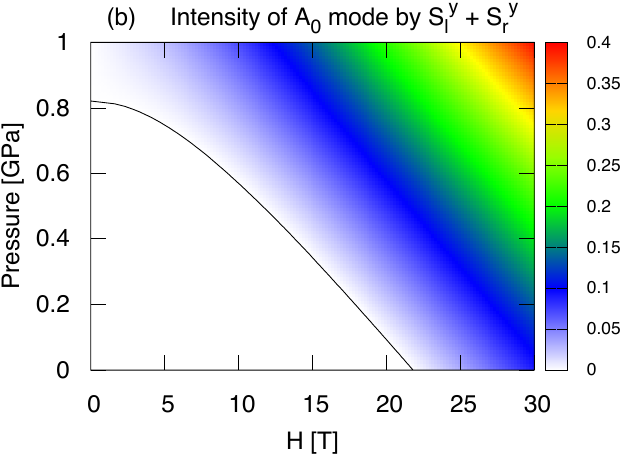}
\caption{
(Color online)
Field dependence of ESR intensity ($|\braket{{\rm ES}|O|{\rm GS}}|^2$) of the A$_0$ mode
by the magnetic channel under various pressures $P$ (GPa) for $\H\parallel [201]$ ($x$ direction).
Here, $|{\rm GS}\rangle$ and $|{\rm ES}\rangle$ are ground and excited states.
$O$ represents an operator causing the magnetical transition.
(a) Intensity for $O=S_l^y+S_r^y$ under various pressures.
This is measurable for $\bH^\omega\parallel y$.
Note that the same result is obtained for $O=S_l^z+S_r^z$ ($\bH^\omega\parallel z$).
(b) Contour map of (a).
For $O=S_l^x+S_r^x$, we remark that the all excitation modes have no intensity for $\bH\parallel x$.
}
\label{fig:intensity-A-mag}
\end{center}
\end{figure}

In the Faraday configuration, where $\bH^\omega$ is in the $yz$ plane,
$S_l^y+S_r^y$ and $S_l^z+S_r^z$ are active, whereas $S_l^x+S_r^x$ is inactive.
In the magnetic channel, a finite intensity only appears in the ordered phase.
Since $S_l^y+S_r^y$ has a finite matrix element between the $|10\rangle$ and $|1\pm1\rangle$ triplet states
[see the corresponding $S_l^x+S_r^x$ in Eq. (\ref{eqn:S+})], only the A$_0$ mode has a finite intensity.
This also holds for $S_l^z+S_r^z$ [$S_l^y+S_r^y$ in Eq. (\ref{eqn:S+})] and it leads to the same result.
The field dependence of the intensity of the A$_0$ mode is shown in Fig. \ref{fig:intensity-A-mag}(a),
which is calculated by the extended spin-wave theory.
The intensity monotonically develops with the field in the ordered phase.
Figure \ref{fig:intensity-A-mag}(b) is the contour map of the intensity.
Thus, the magnetic channel does not contribute to the intensity of the A$_+$ mode even in the ordered phase.

In the presence of magnetic anisotropies, on the other hand, the $|10\rangle$ and $|1\pm1\rangle$ triplet states can be hybridized.
In this case, the magnetic channel can contribute to the intensity of the A$_+$ mode,
however, the main intensity is expected to be from the electric channel owing to the weak magnetic anisotropy.

%%%%%%%%%%%%%%%%%%%%%%%%%%%%%%%%%%%%%%%%%%%%%%%%%%%%%%%%%%%%%%%%%%%%%%%%
%%%%%%%%%%%%%%%%%%%%%%%%%%%%%%%%%%%%%%%%%%%%%%%%%%%%%%%%%%%%%%%%%%%%%%%%

%%%%%%%%%%%%%%%%%%%%%%%%%%%%%%%%%%%%%%%%%%%%%%%%%%%%%%%%%%%%%%%%%%%%%%%%
\subsection{Electric Channel}
%%%%%%%%%%%%%%%%%%%%%%%%%%%%%%%%%%%%%%%%%%%%%%%%%%%%%%%%%%%%%%%%%%%%%%%%

\begin{figure}
\begin{center}
\includegraphics[width=6.25cm]{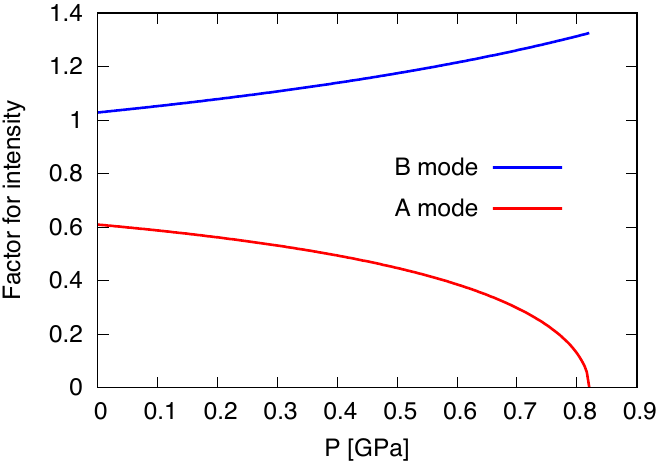}
\caption{
(Color online)
Pressure dependence of the intensity, $(u_\bq+v_\bq)^2=E_\bq/J$, in the disordered phase [see Eq. (\ref{eqn:factor})].
Parameters in Table \ref{table:parameter} were used for the pressure dependences of the exchange interactions.
}
\label{fig:factor}
\end{center}
\end{figure}

We next study the electric channel.
For $P<\Pc$ in the disordered phase, the intensities of all modes (A and B modes) do not dependent on the magnetic field,
as summarized in Table \ref{table:W-linear}.
Intensities of the A and B modes are proportional to $E_0/J$ and $E_{\bk_\rB}/J$, respectively [see Eq. (\ref{eqn:factor})].
They depend on the pressure through the exchange interactions.
We show the pressure dependence of the intensity in Fig. \ref{fig:factor}.
We can see that the intensity of the A mode monotonically decreases with pressure and vanishes at the critical pressure,
while that of the B mode gradually increases.

%%%%%%%%%%%%%%%%%%%%%%%%%%%%%%%%%%%%%%%%%%%%%%%%%%%%%%%%%%%%%%%%%%%%%%%%
%%%%%%%%%%%%%%%%%%%%%%%%%%%%%%%%%%%%%%%%%%%%%%%%%%%%%%%%%%%%%%%%%%%%%%%%

%%%%%%%%%%%%%%%%%%%%%%%%%%%%%%%%%%%%%%%%%%%%%%%%%%%%%%%%%%%%%%%%%%%%%%%%
\subsubsection{A$_0$ Mode}
%%%%%%%%%%%%%%%%%%%%%%%%%%%%%%%%%%%%%%%%%%%%%%%%%%%%%%%%%%%%%%%%%%%%%%%%

\begin{figure*}
\begin{center}
\includegraphics[width=5.9cm]{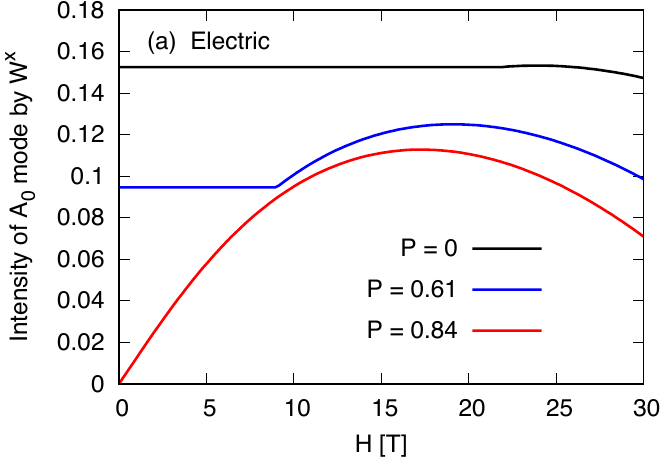}
\includegraphics[width=6cm]{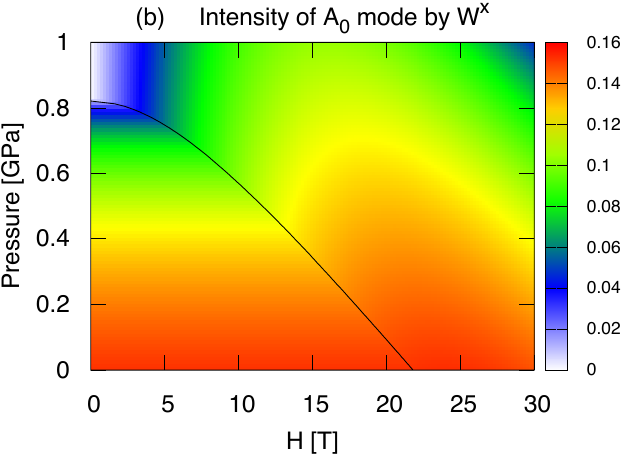}
\includegraphics[width=5.9cm]{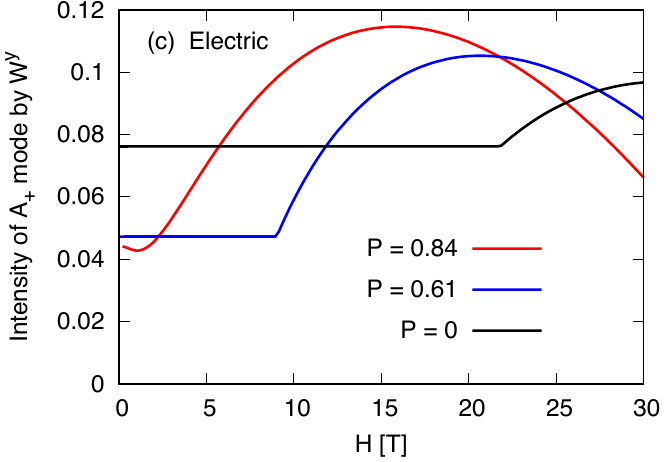}
\includegraphics[width=6cm]{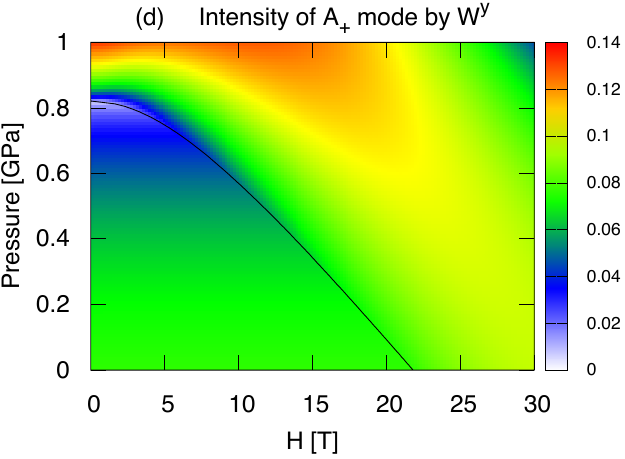}
\includegraphics[width=5.9cm]{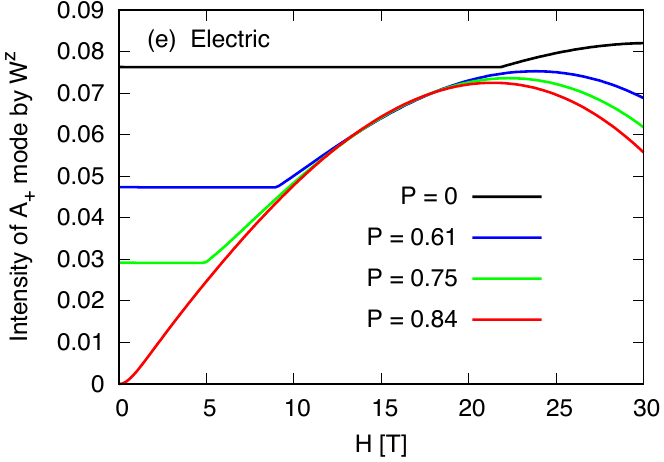}
\includegraphics[width=6cm]{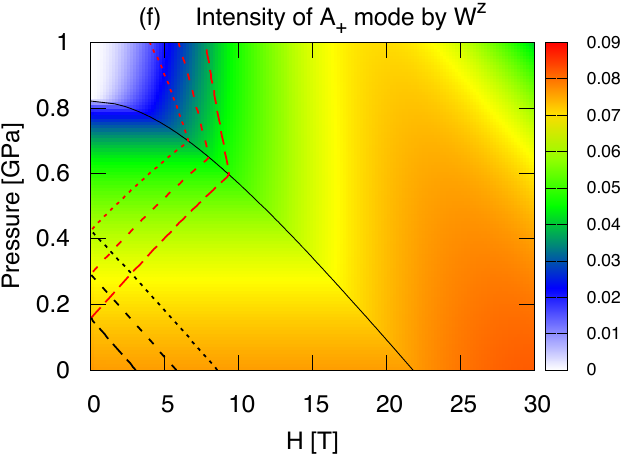}
\caption{
(Color online)
Field dependence of ESR intensity for A modes by the electric channel,
$|\braket{{\rm ES}|O|{\rm GS}}|^2$, under various pressures $P$ (GPa).
Here, $|{\rm GS}\rangle$ and $|{\rm ES}\rangle$ are ground and excited states, respectively.
$O$ represents an operator causing the electrical transition.
For A$_\pm$ modes, the intensity is normalized to be $(u_\bq+v_\bq)^2/8=E_\bq/(8J)$ in the disordered phase.
For A$_0$ mode, it normalized to be $(u_\bq+v_\bq)^2/4=E_\bq/(4J)$ [see Table \ref{table:W-linear}].
(a) A$_0$ mode by the electric channel with $O=W^x$.
This is measurable for $\bE^\omega\parallel x$ via $p^x=C^x_x W^x$
or $\bE^\omega\parallel z$ via $p^z=C^z_x W^x$ [see Eq. (\ref{eqn:P-A})].
(b) Contour map of (a).
The solid line represents the boundary between the disordered and ordered phases.
(c) A$_+$ mode by the electric channel with $O=W^y$.
This is measurable for $\bE^\omega\parallel y$ via $p^y=C^y_y W^y$.
(d) Contour map of (c).
(e) A$_+$ mode by the electric channel with $O=W^z$.
This is measurable for $\bE^\omega\parallel x$ via $p^x=C^x_z W^z$
or $\bE^\omega\parallel z$ via $p^z=C^z_z W^z$.
(f) Contour map of (e).
In the disordered phase, notice that the intensities are the same for $W^z$ and $W^y$.
The red (black) dotted, dashed, and long-dashed lines represent curves 
on which energies of the A$_+$ (A$_-$) mode are 0.381, 0.460, and 0.538 THz, respectively [see Fig. \ref{fig:const-e}].
The lines are plotted for the later convenience.
}
\label{fig:intensity-A-elec}
\end{center}
\end{figure*}

The coefficient tensor for the A modes [see A$_-$, A$_0$, A$_+$ in Fig. \ref{fig:FH}(a)]
is given by $C(+)$ in Eq. (\ref{eqn:C-pm}).
This leads to the following electric dipole for a dimer:
\begin{align}
\begin{pmatrix}
p^x \cr
p^y \cr
p^z
\end{pmatrix}
=
\begin{pmatrix}
C^x_x W^x+C^x_z W^z \cr
C^y_y W^y \cr
C^z_x W^x+C^z_z W^z
\end{pmatrix}.
\label{eqn:P-A}
\end{align}
The matrix elements of the vector spin chirality operators are shown in Eq. (\ref{eqn:v-matrix}).
For $\bH\parallel x$, role of the operators in Eq. (\ref{eqn:v-matrix}) should be replaced as
$(W^x,W^y,W^z) \rightarrow (W^y,W^z,W^x)$, respectively.
For $\bH\parallel x$, $W^x$ [$W^z$ in Eq. (\ref{eqn:v-matrix})]
only has a finite matrix element between the singlet and $|10\rangle$ triplet state.
In the Faraday configuration, $\bE^\omega$ is in the $yz$ plane
and the electric dipole of the $(p^y,p^z)$ components are active in Eq. (\ref{eqn:P-A}).
In Figs. \ref{fig:intensity-A-elec}(a) and \ref{fig:intensity-A-elec}(b),
we show pressure and field dependences of the intensity by $W^x$.
The intensity is suppressed with the increase of the pressure.
In the ordered phase, the suppressed intensity increases with the magnetic field.
We comment that we did not measure the A$_0$ mode in our experiment.

%%%%%%%%%%%%%%%%%%%%%%%%%%%%%%%%%%%%%%%%%%%%%%%%%%%%%%%%%%%%%%%%%%%%%%%%
%%%%%%%%%%%%%%%%%%%%%%%%%%%%%%%%%%%%%%%%%%%%%%%%%%%%%%%%%%%%%%%%%%%%%%%%

%%%%%%%%%%%%%%%%%%%%%%%%%%%%%%%%%%%%%%%%%%%%%%%%%%%%%%%%%%%%%%%%%%%%%%%%
\subsubsection{A$_+$ Mode}
%%%%%%%%%%%%%%%%%%%%%%%%%%%%%%%%%%%%%%%%%%%%%%%%%%%%%%%%%%%%%%%%%%%%%%%%

For $\bH\parallel x$, $W^y$ and $W^z$ [$W^x$ and $W^y$ in Eq. (\ref{eqn:v-matrix})]
have a finite matrix element between the singlet and $|1\pm1\rangle$ triplet states.
Therefore, both $W^y$ and $W^z$ contribute to the intensity of the A$_+$ mode.
In Figs. \ref{fig:intensity-A-elec}(c) and \ref{fig:intensity-A-elec}(e),
we show field dependence of the intensity by $W^y$ and $W^z$, respectively.
Figures \ref{fig:intensity-A-elec}(d) and \ref{fig:intensity-A-elec}(f) are the corresponding contour maps.
In the disordered phase, notice that the intensities are the same for $W^y$ and $W^z$.
For $P<\Pc$, the intensity is suppressed near the critical field and is enhanced in the ordered phase above the critical field.
The enhancement becomes distinct in high-pressure region, since the intensity is strongly suppressed there in the disordered phase.
This can be a possible reason for the observed sharp spectrum of the A$_{++}$ mode in the order phase at $P=0.61$ GPa
(see Fig. \ref{fig:nama-data}).
In addition, the slope of the A$_{++}$ mode is about twice as that of the A$_+$ mode.
This contributes to make the linewidth of A$_{++}$ mode about half that of the A$_+$ mode.

In the Faraday configuration, the $(p^y,p^z)$ components are active in Eq. (\ref{eqn:P-A}).
Since $W^y$ and $W^z$ cause the transition from the ground state to the A$_+$ mode,
$C^y_y$ and $C^z_z$ concern the transition.
$C^z_x$ is related to the intensity of the A$_0$ mode via $p^z=C^z_x W^x$
and this term does not contribute to the intensity of the A$_+$ mode.
In the case of \Tl, it was reported that the ratio of the coefficients is $C^z_z/C^y_y\simeq 124.5/2.5$
by measuring the induced ferroelectricity at the ambient pressure under various field directions in the magnon BEC phase
\cite{Kimura-2020}.
In \K, we also expect a similar ratio, and the main intensity of the A$_+$ mode originates from $P^z=C^z_z W^z$.

For $p^z=C^z_z W^z$, the intensity vanishes in the ordered phase at $H=0$,
as shown in Figs. \ref{fig:intensity-A-elec}(e) and \ref{fig:intensity-A-elec}(f).
Therefore, the A$_+$ mode is hard to detect for $P>\Pc$ at low fields.
On the other hand, the intensity increases with the field
[see Figs. \ref{fig:intensity-A-elec}(e) and \ref{fig:intensity-A-elec}(f)].
This behavior seems to be observed by our measurement,
as we can see in Fig. \ref{fig:higgs}(a) for 0.84 GPa, which is just above the critical pressure ($\Pc=0.82$ GPa).
However, it is not so easy to compare the intensities of the spectra in Fig. \ref{fig:higgs}(a),
since the output of the light source depends on the frequency.
With the same frequency, by contrast, we can compare the intensities.
To examine the field dependence of intensity, we analyze the spectra obtained under the same frequency.
The intensity can be estimated by integrating the peak spectrum over the magnetic field.
The field-integrated intensity is, however, different from the that of the transition probability,
since the latter is defined as an energy-integrated intensity.
There are following relation between the intensities
\cite{note:intensity}:
\begin{align}
I_\omega = g^* I_H.
\label{eqn:int-relation}
\end{align}
Here, $I_\omega$ and $I_H$ are energy- and field-integrated intensities, respectively.
$g^*$ represents a magnetic-field gradient of the excitation energy at the resonance.
We analyze experimental data for $\omega=0.381$, 0.460, and 0.538 THz.

%%%%%%%%%%%%%%%%%%%%%%%%%%%%%%%%%%%%%%%%%%%%%%%%%%%%%%%%%%%%%%%%%%%%%%%%
\begin{figure*}
\begin{center}
\includegraphics[width=6cm]{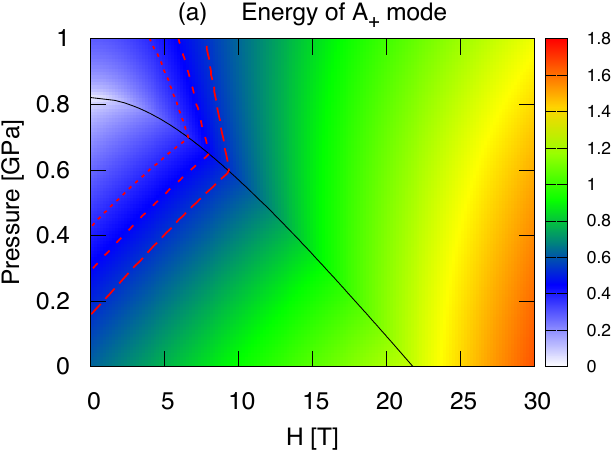}
\includegraphics[width=6cm]{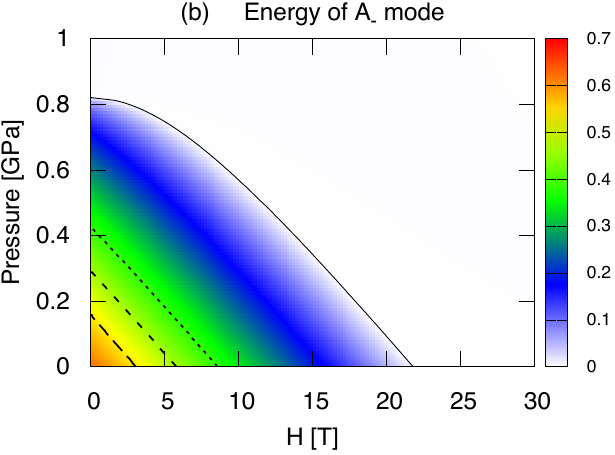}
\caption{
(Color online)
Contour map of excitation energy.
(a) A$_+$ mode.
The red dotted, dashed, and long-dashed lines are for fixed energies of $\omega=0.381$, 0.460, and 0.538 THz, respectively.
The black solid line represents the boundary between the disordered and ordered phases.
(b) A$_-$ mode.
The A$_-$ mode is gapless in the ordered phase.
The black dotted, dashed, and long-dashed lines are for $\omega=0.381$, 0.460, and 0.538 THz, respectively.
}
\label{fig:const-e}
\end{center}
\end{figure*}
%%%%%%%%%%%%%%%%%%%%%%%%%%%%%%%%%%%%%%%%%%%%%%%%%%%%%%%%%%%%%%%%%%%%%%%%

%%%%%%%%%%%%%%%%%%%%%%%%%%%%%%%%%%%%%%%%%%%%%%%%%%%%%%%%%%%%%%%%%%%%%%%%
\begin{figure*}
\begin{center}
\includegraphics[width=12cm]{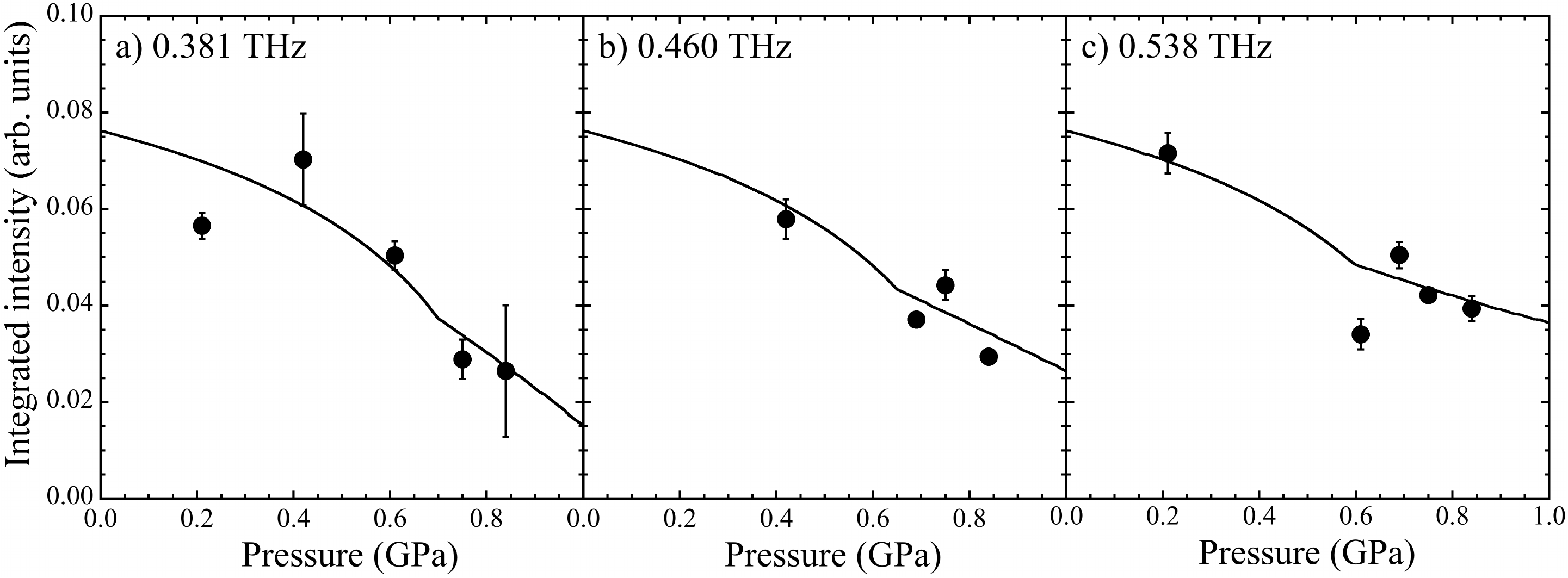}
\caption{
Pressure dependence of the ESR intensity of the A$_+$ mode under fixed excitation energy.
(a) $\omega=0.381$ THz. (b) $\omega=0.460$ THz. (c) $\omega=0.538$ THz.
Solid lines are theoretical curves.
They show a kink at the phase boundary.
The low-pressure region below the kink is in the disordered phase (A$_+$ mode),
whereas the high-pressure region above the kink is in the ordered phase (A$_{++}$ mode).
Points are obtained by integrating the peak spectra over the magnetic field through the relation in Eq. (\ref{eqn:int-relation}).
The data for 0.381 THz at 0.21 GPa is for A$_-$ mode.
Notice that the ESR intensities are the same for the A$_\pm$ modes in the disordered phase.
Therefore, the theoretical curves obtained along the black lines for the A$_-$ mode shown in Fig. \ref{fig:intensity-A-elec}(f)
correspond to the intensity of the A$_-$ mode.
}
\label{fig:intensity-experiment}
\end{center}
\end{figure*}
%%%%%%%%%%%%%%%%%%%%%%%%%%%%%%%%%%%%%%%%%%%%%%%%%%%%%%%%%%%%%%%%%%%%%%%%

First, we explain theoretical results.
In Fig. \ref{fig:const-e}, we show contour maps of excitation energies of A$_\pm$ modes.
There are three (dotted, dashed, and long-dashed) lines
on which excitation energies are constant ($\omega=0.381$, 0.460, and 0.538 THz, respectively).
The three lines are also plotted in Fig. \ref{fig:intensity-A-elec} (f).
The intensities along the lines are the theoretical results.
We show them in Fig. \ref{fig:intensity-experiment} as a function of pressure.
The ESR intensity monotonically decreases with pressure, showing a kink at the phase boundary.
The experimental results are also plotted in Fig. \ref{fig:intensity-experiment}.
With the same $\omega$, we can compare the relative intensities of the experimental results at different pressures.
According to the theoretical curve, the experimental results are plotted with multiplying a common factor for each $\omega$.
Then, we can relatively compare the intensities from different $\omega$.
In Fig. \ref{fig:intensity-experiment}, we can see that the theoretical curve qualitatively explains the experimental result
that the intensity decreases with the increase of pressure.
The important point is that the intensity increases for high frequency.
About the data at 0.84 GPa, the intensity seems to become stronger for $\omega=0.538$
than that for $\omega=0.460$ (see Fig. \ref{fig:intensity-experiment}).
In Fig. \ref{fig:higgs}(a), this means that the intensity at $\omega=0.538$ is stronger than that at $\omega=0.460$.
This indicates that the intensity of the observed excitation mode shown in Fig. \ref{fig:higgs}(c) increases with the magnetic field.
This is consistent with the theoretical result shown in Fig. \ref{fig:intensity-A-elec}(e) for $P=0.84$ GPa,
and supports that the detected excitation mode shown in Fig. \ref{fig:higgs}(c)
is connected to the gapped Higgs amplitude mode at $H=0$.

Similar behavior on the temperature dependence was also reported in \K.
At the ambient pressure, it was reported that the excitation energy of the A$_-$ mode and its intensity
simultaneously decrease with lowering temperature
\cite{Tanaka-1998}.
This can be understood as follows.
It is known that the effective interdimer interactions are reduced when the temperature is increased in the disordered phase,
since thermally excited triplet states disturb their interdimer hoppings
\cite{Ruegg-2005}.
Oppositely, lowering temperature enhances the effective interdimer interactions,
and this has a similar effect of increasing pressure
\cite{Merchant-2014}.
The intensity of the A$_-$ mode then decreases when the temperature is lowered.
Thus, the electric dipole $p^z=C^z_z W^z$ can explain both the pressure and temperature dependences of the ESR intensity.

Another characteristic point of the observed spectrum in Fig. \ref{fig:higgs}(a) is that the linewidth is broadened
compared to that of the A$_{++}$ mode shown in Fig. \ref{fig:nama-data}(b).
This is consistent with the broadened linewidth of the Higgs amplitude mode
observed by inelastic neutron scattering \cite{Ruegg-2008} and Raman scattering \cite{Kuroe-2012} measurements in \Tl.
It is known that the Higgs amplitude mode spontaneously decays into a pair of Nambu--Goldstone modes
and the linewidth is proportional to the excitation gap of the Higgs amplitude mode for three dimensional systems
\cite{Kulik-2011}.
By taking into account the interaction between magnons, the linewidth can be obtained within the extended spin-wave theory
by calculating the transition probability according to the decay process.
Since it involves integral over momentum space and requires time to complete the numerical calculation, this is left as a future work.

Under a finite magnetic field in the ordered phase,
the A$_+$ mode has both longitudinal and transverse fluctuations of the ordered moment.
In this sense, it is not a pure amplitude mode under the field.
As discussed above, the pure Higgs amplitude mode at $H=0$ is hard to detect by ESR in \K~because of losing the intensity.
Thus, the observed signals under the field at 0.84 and 0.94 GPa in Fig. \ref{fig:higgs}
are understood as detecting the field-evolved Higgs amplitude mode.
The observed field dependences of excitation energy and ESR intensity are consistent with the theory.

%%%%%%%%%%%%%%%%%%%%%%%%%%%%%%%%%%%%%%%%%%%%%%%%%%%%%%%%%%%%%%%%%%%%%%%%
\subsubsection{A$_-$ Mode}
%%%%%%%%%%%%%%%%%%%%%%%%%%%%%%%%%%%%%%%%%%%%%%%%%%%%%%%%%%%%%%%%%%%%%%%%

As summarized in Table \ref{table:W-linear}, the intensity of the A$_-$ mode
is exactly the same as that of the A$_+$ mode in the disordered phase.
The energy of the A$_-$ mode decreases with field and becomes soft at the critical field [see Fig. \ref{fig:const-e}(b)].
For isotropic exchange interactions, it stays at zero energy in the ordered phase (Nambu--Goldstone mode).
Therefore, we do not consider the intensity of the A$_-$ mode in the ordered phase.

In the presence of magnetic anisotropies, the energy of the A$_-$ mode acquires an excitation gap in the ordered phase
as in \Tl~\cite{Glazkov-2004}.
In this case, both the magnetic and electric channels contribute to the intensity of the A$_-$ mode
and this leads to a nonreciprocal directional dichroism, as found in \Tl~\cite{Kimura-2020}.

%%%%%%%%%%%%%%%%%%%%%%%%%%%%%%%%%%%%%%%%%%%%%%%%%%%%%%%%%%%%%%%%%%%%%%%%
\subsubsection{B Modes}
%%%%%%%%%%%%%%%%%%%%%%%%%%%%%%%%%%%%%%%%%%%%%%%%%%%%%%%%%%%%%%%%%%%%%%%%

%%%%%%%%%%%%%%%%%%%%%%%%%%%%%%%%%%%%%%%%%%%%%%%%%%%%%%%%%%%%%%%%%%%%%%%%
\begin{figure*}
\begin{center}
\includegraphics[width=6cm]{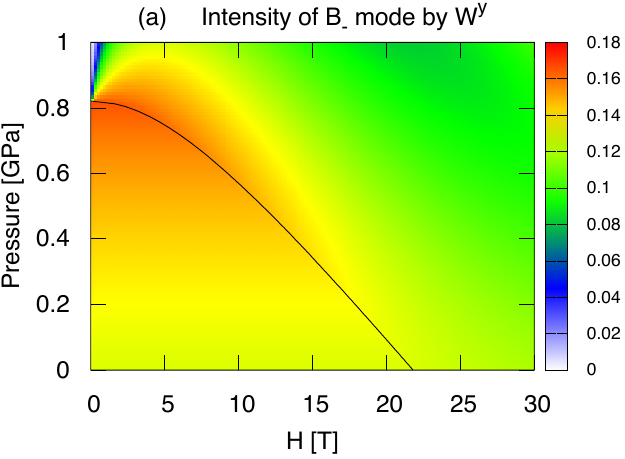}
\includegraphics[width=6cm]{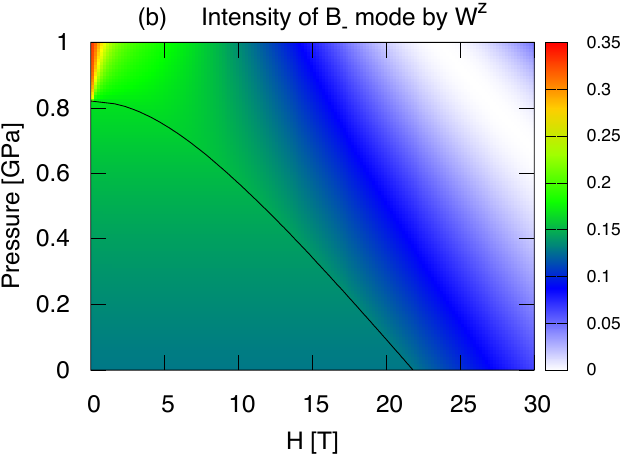}
\includegraphics[width=6cm]{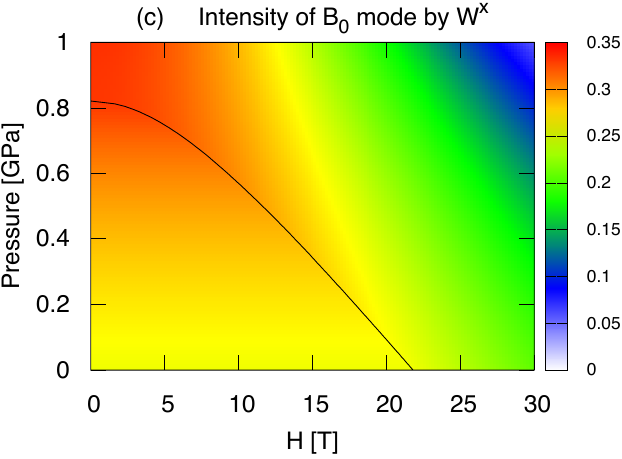} \\
\includegraphics[width=6cm]{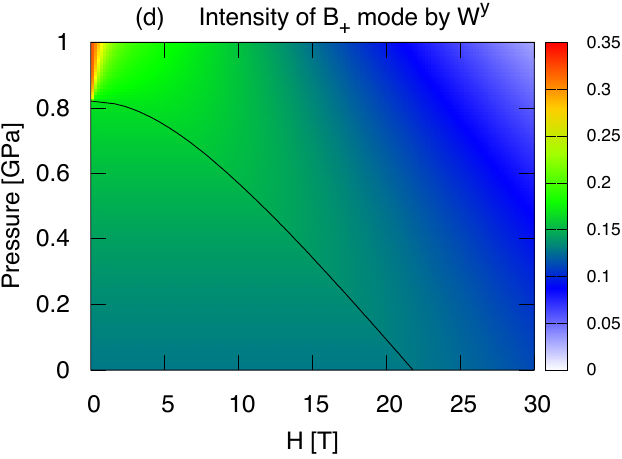}
\includegraphics[width=6cm]{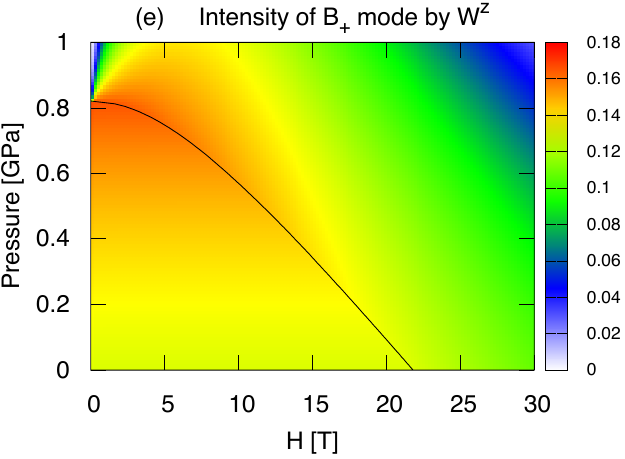}
\caption{
(Color online)
Contour map of pressure and field dependences of the ESR intensity for the B mode.
The solid line represents the phase boundary.
(a) B$_-$ mode by $W^y$.
This is measurable for $\bE^\omega \parallel x$ via $p^x=C^x_y W^y$
or $\bE^\omega \parallel z$ via $p^z=C^z_y W^y$ [see Eq. (\ref{eqn:P-B})].
(b) B$_-$ mode by $W^z$.
This is measurable for $\bE^\omega \parallel y$ via $p^y=C^y_z W^z$.
(c) B$_0$ mode by $W^x$.
This is measurable for $\bE^\omega \parallel y$ via $p^y=C^y_x W^x$.
(d) B$_+$ mode by $W^y$.
This is measurable for $\bE^\omega \parallel x$ via $p^x=C^x_y W^y$
or $\bE^\omega \parallel z$ via $p^z=C^z_y W^y$.
(e) B$_+$ mode by $W^z$.
This is measurable for $\bE^\omega \parallel y$ via $p^y=C^y_z W^z$.
In the disordered phase, the intensity by $W^z$ is the same as that by $W^y$.
}
\label{fig:intensity-3d}
\end{center}
\end{figure*}
%%%%%%%%%%%%%%%%%%%%%%%%%%%%%%%%%%%%%%%%%%%%%%%%%%%%%%%%%%%%%%%%%%%%%%%%

The coefficient tensor for the B modes [see B$_-$, B$_0$, B$_+$ in Fig. \ref{fig:FH}(b)] is given by $C(-)$ in Eq. (\ref{eqn:C-pm}).
This leads to the following electric dipole:
\begin{align}
\begin{pmatrix}
p^x \cr
p^y \cr
p^z
\end{pmatrix}
=
\begin{pmatrix}
C^x_y W^y \cr
C^y_x W^x + C^y_z W^z \cr
C^z_y W^y
\end{pmatrix}.
\label{eqn:P-B}
\end{align}
In Fig. \ref{fig:intensity-3d}, we show contour maps of the ESR intensity for the B mode.
Since this electric dipole is related to the excitation of the B mode,
notice that the dipole is staggered with respect to the A and B sublattices.
In \Tl, the values of $C^\alpha_\beta$ for the A mode in Eq. (\ref{eqn:P-A}) were estimated
by measuring the induced ferroelectricity
\cite{Kimura-2020}.
Since the electric dipole is staggered for the B mode,
$C^\alpha_\beta$ in Eq. (\ref{eqn:P-B}) are difficult to know with static measurements.
On the other hand, we can probe them by measuring the ESR intensity of the B mode.
$C^\alpha_\beta$ for the B mode was examined by ESR in \K~\cite{Kimura-2018},
however, the detail is not yet known.
Our analysis is useful for further investigations for the B mode.

%%%%%%%%%%%%%%%%%%%%%%%%%%%%%%%%%%%%%%%%%%%%%%%%%%%%%%%%%%%%%%%%%%%%%%%%
\section{Summary}
%%%%%%%%%%%%%%%%%%%%%%%%%%%%%%%%%%%%%%%%%%%%%%%%%%%%%%%%%%%%%%%%%%%%%%%%

The high-pressure ESR measurements were performed in \K.
In the disordered phase below the critical pressure ($P<\Pc$),
we confirmed that the excitation gap decreases with pressure
and observed the behavior of the gap closing toward the critical pressure.
The observed pressure dependence of the gap is in good agreement with that
previously reported by inelastic neutron scattering and magnetization measurements (see Fig. \ref{fig:gapP}).
We succeeded in extracting the coefficients for the linear pressure dependence of the exchange interactions
from the observed pressure dependences of the excitation energy of the A and B modes
located at $\bk_\rA=(0,0,0)$ and $\bk_\rB=(0,0,2\pi)$, respectively.

For a spin dimer possessing the inversion center, as in \K~and \Tl, DM interaction is excluded in the intradimer interaction.
The direct transition from the singlet to triplet states is forbidden in the conventional magnetic channel.
We studied ESR intensity on the basis of the vector spin chirality,
which plays the role of an electric dipole operator in a dimer.
The intensity is expressed with an analytic form in the disordered phase.
We summarized the results in Table \ref{table:W-linear}, which is useful for future ESR measurements in spin dimer systems.
We found that the intensity decreases with pressure and vanishes at the critical pressure.
In the pressure-induced ordered phase ($P>\Pc$),
the intensity is zero at $H=0$ and monotonically increases with the magnetic field
[see Figs. \ref{fig:intensity-A-elec}(e) and \ref{fig:intensity-A-elec}(f)].

We observed the ESR spectra of the A$_+$ mode in the ordered phase at 0.84 GPa ($>\Pc$).
The characteristic points are as follows.
(i) The $g$-factor is almost twice as that in the disordered phase.
The value is similar to that of the A$_{++}$ mode observed in the field-induced ordered phase.
(ii) The intensity of the A$_+$ mode develops with the magnetic field
[see Figs. \ref{fig:higgs}(a) and \ref{fig:intensity-experiment}].
(iii) The spectrum is broad and the intensity is weak compared to those of the A$_{++}$ mode at 0.75 GPa ($<\Pc$).
These features coincide with those of the Higgs amplitude mode.
Besides, the second feature (ii) indicates that the observed ESR spectrum is the electric origin.
We carried out the analysis based on the extended spin-wave theory,
using the extracted pressure dependence of the exchange interactions.
It was found that the observed spectra above $\Pc$ are explained by the theoretical curves
for the Higgs amplitude mode under the magnetic field [see Figs. \ref{fig:higgs}(c) and \ref{fig:higgs}(d)].
The observed ESR intensity was also consistent with the theory [see Fig. \ref{fig:intensity-experiment}].
With these findings, we conclude that the Higgs amplitude mode was detected by ESR for the first time.

In \Tl, it is known that the extended spin-wave theory (or bond-operator formulation) quantitatively explains
the magnetization curve, inelastic neutron spectra
\cite{Ruegg-2008,Matsumoto-2004,Matsumoto-2002},
and ESR intensity
\cite{Kimura-2018,Kimura-2020}.
In \K, the theory also explains the magnetization curve and inelastic neutron spectra in the disordered phase
\cite{Matsumoto-2004,Matsumoto-2002}.
In the present paper, we demonstrated that the theory also works in the field- and pressure-induced ordered phases of \K.
Thus, the excited states in both compounds are well described by the extended spin-wave theory.
The main difference between the two compounds is that the interdimer interactions are strong in \Tl,
and \Tl~is close to the quantum critical point under the ambient pressure.
Under pressures, on the other hand, the interdimer interactions are enhanced
and the both compounds are similarly treated by the theory.

The high-pressure ESR measurement is developing and has many potentials to apply in various fields
\cite{Sakurai-2015,Sakurai-2018-1}.
In addition to the inelastic neutron scattering \cite{Ruegg-2008}
and Raman scattering \cite{Kuroe-2012,Matsumoto-2008} measurements,
the high-pressure ESR provides a new way to detect the Higgs amplitude mode
via the electric field component of the microwave.

%%%%%%%%%%%%%%%%%%%%%%%%%%%%%%%%%%%%%%%%%%%%%%%%%%%%%%%%%%%%%%%%%%%%%%%%
\begin{acknowledgements}
The authors express their sincere thanks to S. Kimura for fruitful discussions on ESR by vector spin chirality.
This work was supported by JSPS KAKENHI Grant numbers 19K03746, 19H00648, and 17K05516.
\end{acknowledgements}
%%%%%%%%%%%%%%%%%%%%%%%%%%%%%%%%%%%%%%%%%%%%%%%%%%%%%%%%%%%%%%%%%%%%%%%%

%%%%%%%%%%%%%%%%%%%%%%%%%%%%%%%%%%%%%%%%%%%%%%%%%%%%%%%%%%%%%%%%%%%%%%%%
\appendix
%%%%%%%%%%%%%%%%%%%%%%%%%%%%%%%%%%%%%%%%%%%%%%%%%%%%%%%%%%%%%%%%%%%%%%%%

%%%%%%%%%%%%%%%%%%%%%%%%%%%%%%%%%%%%%%%%%%%%%%%%%%%%%%%%%%%%%%%%%%%%%%%%
\section{Extended Spin-Wave Theory}
\label{sec:extended}
%%%%%%%%%%%%%%%%%%%%%%%%%%%%%%%%%%%%%%%%%%%%%%%%%%%%%%%%%%%%%%%%%%%%%%%%

We apply the extended spin-wave theory, or generalized Holstein--Primakoff theory
\cite{Papanicolaou-1984,Onufrieva-1985,Joshi-1999,Shiina-2003,Shiina-2004},
to describe the magnetic excitations in \K.
The theory is equivalent to the harmonic bond-operator formulation
\cite{Matsumoto-2004,Sommer-2001,Matsumoto-2002,Sachdev-1990,Penc-2012,Rhomhanyi-2012}.
In the bond-operator formulation, a unitary transformation for the spin operator was analytically introduced
to describe the magnetically ordered phase.
In the analytic transformation, many terms were involved in the Hamiltonian for excited states
\cite{Matsumoto-2004,Matsumoto-2002,Sommer-2001},
and the formulation was not so easy to apply to complicated systems.

Within the extended spin-wave theory, in contrast, we can avoid the annoying analytic unitary transformation
\cite{Shiina-2003,Shiina-2004}.
The unitary transformation is numerically performed by solving a mean-field problem.
The mean-field solution is used to calculate matrix elements of the Hamiltonian for the spin-wave excitation.
Excitation modes are obtained by diagonalizing the Hamiltonian.
The numerical procedure is the same in both simple and complicated systems.
Thus, the extended spin-wave theory demonstrates power in analyzing magnetic excitations in complicated systems,
such as spin dimer system \Tl~with magnetic anisotropies
\cite{Ruegg-2008,Kimura-2020,Matsumoto-2008-DM}
and \NH~with inequivalent dimer sites
\cite{Matsumoto-2014}.
In these cases, the theory quantitatively succeeded in explaining the results observed by inelastic neutron scattering and ESR.
Since the theory is not yet widely known, we present details of the theory in this paper.
The formulation is given in a general form for magnetic anisotropies
to show that the extended spin-wave theory has no difficulty to apply to complicated systems.
%In the following, the formulation is given according to Ref. \ref{ref:Shiina-2003}.
In the following, the formulation is given according to Ref. 47.

%%%%%%%%%%%%%%%%%%%%%%%%%%%%%%%%%%%%%%%%%%%%%%%%%%%%%%%%%%%%%%%%%%%%%%%%
\subsection{Hamiltonian}
%%%%%%%%%%%%%%%%%%%%%%%%%%%%%%%%%%%%%%%%%%%%%%%%%%%%%%%%%%%%%%%%%%%%%%%%

We consider the Hamiltonian for \K~in the following general form:
\begin{align}
\H = \sum_i \sum_\mu \H_{\rm intra}(i,\mu) + \sum_{\langle i\mu j\mu' \rangle} \H_{\rm inter}(i,\mu,j,\mu').
\label{eqn:H}
\end{align}
Here, $\H_{\rm intra}(i,\mu)$ is for intradimer interaction on the $\mu(={\rm A},{\rm B})$ sublattice in the $i$th unit cell.
It is expressed as
\begin{align}
\H_{\rm intra}(i,\mu)
&= \sum_{\alpha\beta=x,y,z} S_{i\mu l}^\alpha J_\mu^{\alpha\beta} S_{i\mu r}^\beta \cr
&~~~
- \sum_{\gamma=l,r} \sum_{\alpha\beta=x,y,z} H^\alpha g_{\mu\gamma}^{\alpha\beta} \mu_{\rm B} S_{i\mu\gamma}^\beta.
\label{eqn:H-intra}
\end{align}
Here, $J_\mu^{\alpha\beta}$ represents a tensor for the intradimer interaction on the $\mu$ sublattice.
It connects the $\alpha$ and $\beta$ components of the spin operators
on the left ($S_{i\mu l}^\alpha$) and right ($S_{i\mu r}^\beta$) sides of a dimer
on the $\mu$ sublattice in the $i$th unit cell,
where the left and right side of a dimer are denoted by subscripts $l$ and $r$, respectively.
$g_\gamma^{\alpha\beta}$ is the $g$-tensor on the $\gamma(=l,r)$ side of a dimer.
Note that the tensors on the ${\rm A}$ and ${\rm B}$ sublattices are not independent.
They are related by symmetry transformations allowed in the space group to which \K~belongs.

In Eq. (\ref{eqn:H}), $\H_{\rm inter}(i,\mu,j,\mu')$ represents the interdimer interaction
between the $\mu$ and $\mu'$ sublattices in the $i$th and $j$th unit cell.
It is expressed as
\begin{align}
\H_{\rm inter}(i,\mu,j,\mu')
= \sum_{\gamma,\gamma'=l,r} \sum_{\alpha\beta}
S_{i\mu\gamma}^\alpha J_{i\mu\gamma,j\mu'\gamma'}^{\alpha\beta} S_{j\mu'\gamma'}^\beta.
\end{align}
Here, $J_{i\mu\gamma,j\mu'\gamma'}^{\alpha\beta}$ are tensors for the interdimer interactions.
The dominant interaction paths are shown in Fig. \ref{fig:interaction} with $J_1$, $J_2$, and $J_3$.
The summation $\sum_{\langle i\mu j\mu' \rangle}$ in Eq. (\ref{eqn:H}) is taken over the spin pairs connected by the interactions.

In the absence of the magnetic anisotropies, or when they are small and are not important,
the tensors for the anisotropies will be treated as
\begin{align}
\begin{aligned}
&g_{\mu\gamma}^{\alpha\beta}=g\delta_{\alpha\beta}, \cr
&J_\mu^{\alpha\beta}=J\delta_{\alpha\beta}, \cr
&J_{i\mu\gamma,j\mu'\gamma'}^{\alpha\beta}=J_{i\mu\gamma,j\mu'\gamma'} \delta_{\alpha\beta}.
\end{aligned}
\label{eqn:g-isotropic}
\end{align}
Here, $\delta_{\alpha\beta}$ is the Kronecker delta.
$g$, $J$, and $J_{i\mu\gamma,j\mu'\gamma'}$ represent isotropic $g$-factor, intradimer, and interdimer interactions, respectively.

%%%%%%%%%%%%%%%%%%%%%%%%%%%%%%%%%%%%%%%%%%%%%%%%%%%%%%%%%%%%%%%%%%%%%%%%
\subsection{Mean-Field Solution}
%%%%%%%%%%%%%%%%%%%%%%%%%%%%%%%%%%%%%%%%%%%%%%%%%%%%%%%%%%%%%%%%%%%%%%%%

The mean-field Hamiltonian for the dimer on the $\mu$ sublattice in the $i$th unit cell is expressed as
\begin{align}
\H_{\rm MF}(i,\mu) &= \H_{\rm intra}(i,\mu) \cr
&
+ \sum_{j\mu'} \sum_{\gamma\gamma'} \sum_{\alpha\beta}
 S_{i\mu\gamma}^\alpha J_{i\mu\gamma,j\mu'\gamma'}^{\alpha\beta} \braket{S_{j\mu'\gamma'}^\beta}.
\label{eqn:H-mf}
\end{align}
Here, the summation $\sum_{j\mu'}$ and $\sum_{\gamma\gamma'}$ are taken over the neighboring spins
connected to $S_{i\mu\gamma}^\alpha$ by the interdimer interactions.
Since a dimer is composed of two $S=1/2$ spins in \K,
the Hamiltonian is expressed in a $4\times 4$ matrix form and it is solved by diagonalization.
The energy eigenstates are expressed with $|m\rangle$ ($m=0,1,2,3$).
Here, $m=0$ represents the ground state, whereas $m=1,2,3$ are for excited states.
In Eq. (\ref{eqn:H-mf}), the expectation value is given by
$\braket{S_{j\mu'\gamma'}^\beta} = \braket{0 |S_{j\mu'\gamma'}^\beta| 0}$
at sufficiently low temperatures.
Since $\H_{\rm MF}(i,\mu)$ is independent on $i$ for the unit cell,
we have two mean-field Hamiltonians, reflecting the two ($\mu={\rm A},{\rm B})$ sublattices.
We iteratively solve the mean-field problem until the expectation values converge.
In the disordered phase, the expectation value of the spin operator vanishes.
In the ordered phase, it takes a finite value.

%%%%%%%%%%%%%%%%%%%%%%%%%%%%%%%%%%%%%%%%%%%%%%%%%%%%%%%%%%%%%%%%%%%%%%%%
\subsection{Magnetic Excitation}
%%%%%%%%%%%%%%%%%%%%%%%%%%%%%%%%%%%%%%%%%%%%%%%%%%%%%%%%%%%%%%%%%%%%%%%%

The extended spin-wave theory is constructed on the basis of the mean-field solution, as in the linear spin-wave theory.
The intradimer part of the Hamiltonian can be expressed as
\begin{align}
\H_{\rm intra}(i,\mu) = \sum_{m,n=0}^3 |i\mu m\rangle \langle i\mu m| \H_{\rm intra}(i,\mu) | i\mu n \rangle \langle i\mu n|.
\end{align}
Here, $|i\mu m\rangle$ represents the $m$th eigenstate of the mean-field Hamiltonian on the $\mu$ sublattice in the $i$th unit cell.
At each dimer site, we introduce the following bosons which create the eigenstates out of the vacuum state
\cite{Shiina-2003}:
\begin{align}
a_{i\mu m}^\dagger |{\rm vac}\rangle = |i\mu m\rangle.
\end{align}
Here, we introduced different Bose operators on the A and B sublattices.
This means that we adopt the reduced zone scheme for the Brillouin zone, reflecting the two sublattices in the unit cell.
The bosons are subjected to the following local constraint:
\begin{align}
\sum_{m=0}^3 a_{i\mu m}^\dagger a_{i\mu m} = 1.
\label{eqn:constraint-a}
\end{align}
$\H_{\rm intra}(i,\mu)$ is then expressed with the bosons as
\begin{align}
\H_{\rm intra}(i,\mu) = \sum_{m,n=0}^3 \H_{mn}^{\rm intra}(\mu) a_{i\mu m}^\dagger a_{i\mu n},
\label{eqn:H-i}
\end{align}
where $\H_{mn}^{\rm intra}(\mu)$ is the matrix element defined by
\begin{align}
\H_{mn}^{\rm intra}(\mu) = \langle i\mu m| \H_{\rm intra}(i,\mu) | i\mu n \rangle.
\label{eqn:H-mn}
\end{align}
Similarly, the spin operator is expressed as
\begin{align}
S_{i\mu\gamma}^\alpha &= \sum_{m,n=0}^3 S_{mn}^\alpha(\mu,\gamma) a_{i\mu m}^\dagger a_{i\mu m} \cr
&=S_{00}^\alpha(\mu,\gamma) a_{i\mu 0}^\dagger a_{i\mu 0}
+ \sum_{m=1}^3 S_{m0}^\alpha(\mu,\gamma) a_{i\mu m}^\dagger a_{i\mu 0} \cr
&~~~
+ \sum_{m=1}^3 S_{0m}^\alpha(\mu,\gamma) a_{i\mu 0}^\dagger a_{i\mu m} \cr
&~~~
+ \sum_{m,n=1}^3 S_{mn}^\alpha(\mu,\gamma) a_{i\mu m}^\dagger a_{i\mu m}.
\label{eqn:S-i}
\end{align}
with
\begin{align}
S_{mn}^\alpha(\mu,\gamma) = \langle i\mu m| S_{i\mu\gamma}^\alpha | i\mu n \rangle.
\label{eqn:S-mn}
\end{align}
Notice that the matrix elements $\H_{mn}^{\rm intra}(\mu)$ and $S_{mn}^\alpha(\mu,\gamma)$
in Eqs. (\ref{eqn:H-mn}) and (\ref{eqn:S-mn}), respectively, are independent on the site index $i$.

Next, we replace the $a_{i\mu 0}$ operator for the mean-field ground state with use of the local constraint as
\begin{align}
a_{i\mu 0}^\dagger a_{i\mu 0} = M - \sum_{m=1}^3 a_{i\mu m}^\dagger a_{i\mu m}.
\label{eqn:holstein-0}
\end{align}
Here, we introduced $M$ as an expansion parameter for the theory, which we can finally put $M=1$ in the formulation.
For the second and third terms in Eq. (\ref{eqn:S-i}), it is convenient to introduce the following generalized Holstein--Primakoff method
\cite{Papanicolaou-1984,Onufrieva-1985,Joshi-1999,Shiina-2003}:
\begin{equation}
\begin{aligned}
&a_{i\mu m}^\dagger a_{i\mu 0} \rightarrow a_{i\mu m} \left( M - \sum_{m=1}^3 a_{i\mu m}^\dagger a_{i\mu m} \right)^{1/2}, \cr
&a_{i\mu 0}^\dagger a_{i\mu m} \rightarrow \left( M - \sum_{m=1}^3 a_{i\mu m}^\dagger a_{i\mu m} \right)^{1/2} a_{i\mu m}.
\end{aligned}
\label{eqn:holstein}
\end{equation}
Expanding the square root, we can express the spin operator in Eq. (\ref{eqn:S-i}) in powers of $M^{-1}$.
We substitute Eqs. (\ref{eqn:holstein-0}) and (\ref{eqn:holstein}) into Eqs. (\ref{eqn:H-i}) and (\ref{eqn:S-i})
and eliminate the $a_{i\mu 0}$ and $a_{i\mu 0}^\dagger$ operators.
$\H_{\rm intra}(i,\mu)$ and $S_{i\mu\gamma}^\alpha$ are expressed with the bosons for the excited states ($m \ge 1$).
The Hamiltonian [Eq. (\ref{eqn:H})] is then written as
\cite{Shiina-2003}
\begin{align}
\H = M^2 \sum_{n=0}^\infty M^{-\frac{n}{2}} \H^{(n)}.
\end{align}
Here, $\H^{(n)}$ represents $n$-boson Hamiltonian.
$\H^{(0)}$ is a c-number term that corresponds to energy of the the mean-field ground state.
$\H^{(1)}$ is the first-order term of the bosons for the excited states.
It vanishes with the self-consistently determined mean-field solution.
$\H^{(2)}$ is the second-order (harmonic) term.
Since bosons are dilute at low temperatures, we neglect higher order terms of bosons and put $M=1$.

The Hamiltonian is then expressed in the following form:
\begin{align}
\H = \H_{\rm local} + \H_{\rm nonlocal}.
\end{align}
Here, $\H_{\rm local}$ and $\H_{\rm nonlocal}$ are for local and nonlocal parts, respectively.
They are given by
\begin{align}
&\H_{\rm local} = \sum_i \sum_{m,n=1}^3 \sum_{\mu={\rm A,B}} \Lambda_{mn}^\mu a_{i\mu m}^\dagger a_{i\mu n},
\label{eqn:H-total} \\
&\H_{\rm nonlocal} = \sum_{i\mu j\mu'} \sum_{m,n=1}^3
\left[
  \Pi_{mn}^{\mu\mu'}(i,j) a_{i\mu m}^\dagger a_{j\mu' n} \right. \cr
&~~~~~~~~~~~~~~~~~~~~~~~~\left.  
+ \Delta_{mn}^{\mu\mu'}(i,j) a_{i\mu m}^\dagger a_{j\mu' n}^\dagger + {\rm h. c.}
\right],
\nonumber
\end{align}
with
\begin{align}
&\Lambda_{mn}^\mu = \H_{mn}^{\rm intra}(\mu) - \H_{00}^{\rm intra}(\mu)\delta_{mn} \cr
&~~~~~~~
+ \sum_{j\mu'\gamma\gamma'} \sum_{\alpha\beta} J_{i\mu\gamma,j\mu'\gamma'}^{\alpha\beta}
    \left[ S_{mn}^\alpha(\mu,\gamma) - S_{00}^\alpha(\mu,\gamma) \delta_{mn} \right] \cr
&~~~~~~~~~~~~~~~~~~~~~
\times S_{00}^\beta(\mu',\gamma'),
\label{eqn:Lambda-Pi} \\
&\Pi_{mn}^{\mu\mu'}(i,j) = \sum_{\gamma\gamma'} \sum_{\alpha\beta} J_{i\mu\gamma,j\mu'\gamma'}^{\alpha\beta}
   S_{m0}^\alpha(\mu,\gamma) S_{0n}^\beta(\mu',\gamma'), \cr
&\Delta_{mn}^{\mu\mu'}(i,j)
= \sum_{\gamma\gamma'} \sum_{\alpha\beta} J_{i\mu\gamma,j\mu'\gamma'}^{\alpha\beta}
   S_{m0}^\alpha(\mu,\gamma) S_{n0}^\beta(\mu',\gamma').
\nonumber
\end{align}
We next introduce the Fourier-transformation as
\begin{equation}
\begin{aligned}
&a_{i\mu m} = \frac{1}{\sqrt{N/2}} \sum_\bk a_{\bk\mu m} e^{i \bk\cdot\bm{R}_{i\mu}}, \cr
&a_{\bk \mu m} = \frac{1}{\sqrt{N/2}} \sum_i a_{i\mu m} e^{-i \bk\cdot\bm{R}_{i\mu}}.
\end{aligned}
\label{eqn:Fourier-boson}
\end{equation}
Here, $N$ is number of dimer sites, and $N/2$ represents number of unit cell in the sample.
$\bm{R}_{i\mu}$ represents the position of a dimer on the $\mu$ sublattice in the $i$ th unit cell.
Substituting Eq. (\ref{eqn:Fourier-boson}) into Eq. (\ref{eqn:H-total}), we obtain
\begin{align}
\H &= \sum_\bk \sum_{m,n=1}^3 \sum_\mu \Lambda_{mn}^\mu a_{\bk\mu m}^\dagger a_{\bk\mu n} \cr
&+ \sum_\bk \sum_{m,n=1}^3 \sum_{\mu\mu'} \sum_\nu
      \left[ \left( \Pi_{mn}^{\mu\mu'(\nu)} a_{\bk\mu m}^\dagger a_{\bk\mu'n} \right. \right. \cr
&~~~~~~~~~~~~~~~~~~\left. \left.
+ \Delta_{mn}^{\mu\mu'(\nu)} a_{\bk\mu m}^\dagger a_{-\bk\mu'n}^\dagger \right) \gamma_\bk^{(\nu)} + {\rm h. c.} \right].
\label{eqn:H-diag2} \cr
\end{align}
Here, the superscript $(\nu)$ represents different kinds of interdimer interaction path.
The explicit form of $\Lambda_{mn}^\mu$ is given by
\begin{align}
\Lambda_{mn}^{\mu} &= \Lambda_{mn}^{\rm \mu 0} \cr
&+ \sum_{\alpha\beta} 2 J_{3}^{\alpha\beta} \left[ S_{mn}^\alpha(\mu,l) - S_{00}^\alpha(\mu,l) \delta_{mn} \right] S_{00}^\beta(\bar{\mu},r) \cr
&+ \sum_{\alpha\beta} 2 J_{3}^{\alpha\beta} \left[ S_{mn}^\alpha(\mu,r) - S_{00}^\alpha(\mu,r) \delta_{mn} \right] S_{00}^\beta(\bar{\mu},l). \cr
\end{align}
Here, $\bar{\mu}$ takes $\bar{\mu}=({\rm B,A})$ for $\mu=({\rm A,B})$, respectively.
$\Lambda_{mn}^{\mu0}$ is defined by
\begin{align}
\Lambda_{mn}^{\mu0} &= \H_{mn}^{\rm intra}(\mu) - \H_{00}^{\rm intra}(\mu)\delta_{mn} \cr
&+ \sum_{\alpha\beta} (J_1^{\alpha\beta}+J_2^{\alpha\beta})
      \left[ S_{mn}^\alpha(\mu,l) - S_{00}^\alpha(\mu,l) \delta_{mn} \right] S_{00}^\beta(\mu,r) \cr
&+ \sum_{\alpha\beta} (J_1^{\alpha\beta}+J_2^{\alpha\beta})
      \left[ S_{mn}^\alpha(\mu,r) - S_{00}^\alpha(\mu,r) \delta_{mn} \right] S_{00}^\beta(\mu,l). \cr
\label{eqn:rambda}
\end{align}
$\Pi_{mn}^{\rm \mu\mu'(\nu)}$ is given by
\begin{align}
&\Pi_{mn}^{{\rm AA}(1,0,0)} = \sum_{\alpha\beta} J_{1}^{\alpha\beta} S_{m0}^\alpha({\rm A},r) S_{0n}^\beta({\rm A},l), \cr
&\Pi_{mn}^{{\rm BB}(-1,0,0)} = \sum_{\alpha\beta} J_{1}^{\alpha\beta} S_{m0}^\alpha({\rm B},r) S_{0n}^\beta({\rm B},l), \cr
&\Pi_{mn}^{{\rm AA}(2,0,1)} = \sum_{\alpha\beta} J_{2}^{\alpha\beta} S_{m0}^\alpha({\rm A},r) S_{0n}^\beta({\rm A},l),
\label{eqn:pi} \\
&\Pi_{mn}^{{\rm BB}(-2,0,-1)} = \sum_{\alpha\beta} J_{2}^{\alpha\beta} S_{m0}^\alpha({\rm B},r) S_{0n}^\beta({\rm B},l), \cr
&\Pi_{mn}^{{\rm \mu\bar{\mu}}(1,\frac{1}{2},\frac{1}{2})} = \Pi_{mn}^{{\rm \mu\bar{\mu}}(-1,\frac{1}{2},-\frac{1}{2})}
= \sum_{\alpha\beta} J_{3}^{\alpha\beta} S_{m0}^\alpha(\mu,r) S_{0n}^\beta(\bar{\mu},l).
\nonumber
\end{align}
Here, the superscripts
$(1,0,0)$, $(-1,0,0)$, $(2,0,1)$, $(-2,0,-1)$, $(1,\frac{1}{2},\frac{1}{2})$, and $(1,\frac{1}{2},\frac{1}{2})$
represent the path directions of the interdimer interactions.
$\Delta_{mn}^{\mu\mu'(\nu)}$ is given by
\begin{align}
&\Delta_{mn}^{{\rm AA}(1,0,0)} = \sum_{\alpha\beta} J_{1}^{\alpha\beta} S_{m0}^\alpha({\rm A},r) S_{n0}^\beta({\rm A},l), \cr
&\Delta_{mn}^{{\rm BB}(-1,0,0)} = \sum_{\alpha\beta} J_{1}^{\alpha\beta} S_{m0}^\alpha({\rm B},r) S_{n0}^\beta({\rm B},l), \cr
&\Delta_{mn}^{{\rm AA}(2,0,1)} = \sum_{\alpha\beta} J_{2}^{\alpha\beta} S_{m0}^\alpha({\rm A},r) S_{n0}^\beta({\rm A},l),
\label{eqn:Delta} \\
&\Delta_{mn}^{{\rm BB}(-2,0,-1)} = \sum_{\alpha\beta} J_{2}^{\alpha\beta} S_{m0}^\alpha({\rm B},r) S_{n0}^\beta({\rm B},l), \cr
&\Delta_{mn}^{\mu\bar{\mu}(1,\frac{1}{2},\frac{1}{2})} = \Delta_{mn}^{\mu\bar{\mu}(-1,\frac{1}{2},-\frac{1}{2})}
= \sum_{\alpha\beta} J_{3}^{\alpha\beta} S_{m0}^\alpha(\mu,r) S_{n0}^\beta(\bar{\mu},l).
\nonumber
\end{align}
$\gamma_\bk^{(\nu)}$ is given by
\begin{align}
&\gamma_\bk^{(1,0,0)} = e^{-i k_a},~~~~~~~~~~~~~~~
\gamma_\bk^{(-1,0,0)} = e^{i k_a}, \cr
&\gamma_\bk^{(2,0,1)} = e^{-i (2k_a+k_c)},~~~~~~~~
\gamma_\bk^{(-2,0,-1)} = e^{i (2k_a+k_c)}, \cr
&\gamma_\bk^{(1,\frac{1}{2},\frac{1}{2})} = e^{-i (k_a+\frac{k_b}{2}+\frac{k_c}{2})},~~~
\gamma_\bk^{(-1,\frac{1}{2},-\frac{1}{2})} = e^{-i (-k_a+\frac{k_b}{2}-\frac{k_c}{2})}. \cr
\end{align}
Here, the wavevector is expressed as
\begin{align}
\bm{k} = (k_a,k_b,k_c)
\end{align}
in the reciprocal lattice space.

We next introduce the following 6(=3$\times$2)-dimensional transformed vectors:
\begin{equation}
\begin{aligned}
&{\bm{a}_\bk}^{\rm T}
=
\left(
  \begin{matrix}
    \cdots a_{\bk{\rm A}m} \cdots a_{\bk{\rm B}m} \cdots
  \end{matrix}
\right), \cr
&{\bm{a}_{-\bk}^\dagger}^{\rm T}
=
\left(
  \begin{matrix}
    \cdots a_{-\bk{\rm A}m}^\dagger \cdots a_{-\bk{\rm B}m}^\dagger \cdots
  \end{matrix}
\right).
\end{aligned}
\end{equation}
Using these vectors, we introduce the 12-dimensional vector
\begin{align}
\overrightarrow{a}_\bk =
\left(
  \begin{array}{c}
    \bm{a}_\bk \cr
    \bm{a}_{-\bk}^\dagger
  \end{array}
\right).
\end{align}
Then, we obtain the commutation relation
\begin{align}
\H \overrightarrow{a}_\bk - \overrightarrow{a}_\bk \H = - \hat{\epsilon}_\bk \overrightarrow{a}_\bk.
\label{eqn:commutation}
\end{align}
Here, $\hat{\epsilon}_\bk$ is expressed as
\begin{align}
\hat{\epsilon}_\bk =
\left(
  \begin{array}{cc}
    \bm{A}_\bk & \bm{B}_\bk \cr
    - \bm{B}_{-\bk}^* & - \bm{A}_{-\bk}^*
  \end{array}
\right),
\label{eqn:epsilon-k}
\end{align}
with
\begin{align}
\begin{aligned}
&\bm{A}_\bk =
\begin{pmatrix}
    \bm{\Pi}_\bk^{\rm AA} & \bm{\Pi}_\bk^{\rm AB} \cr
    \bm{\Pi}_\bk^{\rm BA} & \bm{\Pi}_\bk^{\rm BB}  \cr
\end{pmatrix}, \cr
&\bm{B}_\bk =
\begin{pmatrix}
    \bm{\Delta}_\bk^{\rm AA} & \bm{\Delta}_\bk^{\rm AB} \cr
    \bm{\Delta}_\bk^{\rm BA} & \bm{\Delta}_\bk^{\rm BB} \cr
\end{pmatrix}.
\end{aligned}
\end{align}
Here, $\bm{\Lambda}^\mu$, $\bm{\Pi}_\bk^{\mu\mu'}$, and $\bm{\Delta}_\bk^{\mu\mu'}$ are $3\times 3$ matrices.
Their matrix elements are given by
\begin{equation}
\begin{aligned}
&\left(\bm{\Pi}_\bk^{\mu\mu'}\right)_{mn} = \delta_{\mu\mu'} \Lambda_{mn}^\mu
+ \sum_\nu \left[ \Pi_{mn}^{\mu\mu'}\gamma_\bk^{(\nu)} + \left( \Pi_{nm}^{\mu'\mu}\gamma_\bk^{(\nu)} \right)^* \right], \cr
&\left(\bm{\Delta}_\bk^{\mu\mu'}\right)_{mn} =
  \Delta_{mn}^{\mu\mu'}\gamma_\bk + \Delta_{nm}^{\mu'\mu}\gamma_{-\bk}.
\end{aligned}
\end{equation}

Next, we assume that the Hamiltonian has the following diagonal form:
\begin{align}
\H = \sum_\bk E_\bk \alpha_\bk^\dagger \alpha_\bk.
\end{align}
Here, $\alpha_\bk$ is a boson for a magnon excitation.
It satisfies the commutation relation
\begin{align}
\H \alpha_\bk - \alpha_\bk \H = - E_\bk \alpha_\bk.
\end{align}
To diagonalize the Hamiltonian in Eq. (\ref{eqn:H-diag2}), we introduce the following Bogoliubov transformation:
\begin{align}
\alpha_\bk = \overrightarrow{X}_\bk^{\rm T} \overrightarrow{a}_\bk.
\label{eqn:Bogoliubov-0}
\end{align}
Here, $\overrightarrow{X}_\bk^{\rm T}$ is a 12-dimensional transposed vector.
Substituting Eq. (\ref{eqn:Bogoliubov-0}) into Eq. (\ref{eqn:commutation}), we obtain the eigenvalue equation
\begin{align}
\hat{\epsilon}_\bk^{\rm T} \overrightarrow{X}_\bk = E_\bk \overrightarrow{X}_\bk.
\label{eqn:eigen}
\end{align}
When we diagonalize the $12\times 12$ matrix $\hat{\epsilon}_\bk^{\rm T}$,
the energy eigenvalues are obtained as pairs of $\pm E_\bk$.
Here, $\pm$ are for particles and holes, respectively.
We take only particle (positive eigenvalue) solutions.
There are six excitation modes in the Brillouin zone reflecting the two dimer sublattices,
since each dimer carries three excited states.

The extended spin-wave theory presented in this section may seem to be complicated, however, it is not.
It is easy to handle in the following way.
Under a given pressure and external magnetic field,
we first iteratively solve the mean-field problem given by Eq. (\ref{eqn:H-mf}) in a $4\times 4$ matrix form.
The mean-field solutions are used to calculate the matrix elements of Eq. (\ref{eqn:epsilon-k}).
Under a fixed wavevector $\bk$, we next solve the eigenvalue problem of Eq. (\ref{eqn:eigen}).
The matrix is described in a $12\times 12$ form, and the numerical diagonalization is easy to perform.
The dispersion relations of the 6 modes are obtained by varying the wavevector $\bk$
in both the disordered and ordered phases.

The extended spin-wave theory was actually used to analyze the pressure-dependence of the excitation modes
in \Tl~through a quantum critical point in the presence of a magnetic anisotropy
\cite{Ruegg-2008}.
In the disordered phase, it was revealed that the triplet excited states split into three modes
due to the anisotropy and that the lowest mode
becomes soft on the quantum critical point ($P=\Pc \simeq 1$ kbar $=0.1$ GPa).
In the ordered phase ($P>\Pc$), the lowest mode changes into a Higgs amplitude mode (longitudinal mode),
acquiring an energy gap that develops with the pressure.
The calculated results are quantitatively consistent with the observed data by inelastic neutron scattering performed under pressures
\cite{Ruegg-2008}.
In this paper, we apply the established extended spin-wave theory to the isostructural compound \K~for ESR.

%%%%%%%%%%%%%%%%%%%%%%%%%%%%%%%%%%%%%%%%%%%%%%%%%%%%%%%%%%%%%%%%%%%%%%%%
\subsection{Intensity of ESR in Ordered Phase}
%%%%%%%%%%%%%%%%%%%%%%%%%%%%%%%%%%%%%%%%%%%%%%%%%%%%%%%%%%%%%%%%%%%%%%%%

The extended spin-wave theory was used to analyze the nonreciprocal directional dichroism
found by ESR in \Tl~\cite{Kimura-2020}.
We introduce the formulation here.

Since magnetic channel can be active to ESR in the ordered phase, we treat both the electromagnetic fields.
The perturbation Hamiltonian for microwave absorption is then given by
\begin{equation}
\begin{aligned}
&\H'(t) = \left( \H'_M + \H'_P \right) e^{-i\omega t}, \cr
&\H'_M = - \bH^\omega \cdot \bM, \cr
&\H'_P = - \bE^\omega \cdot \bP.
\end{aligned}
\end{equation}
with
\begin{equation}
\begin{aligned}
&M^\alpha= \sum_{\beta=x,y,z} g \mu_{\rm B} \sum_i \sum_\mu \sum_\gamma S_{i\mu\gamma}^\alpha~~~(\alpha=x,y,z), \cr
&\bP = \sum_i \sum_\mu \bp_{i\mu}.
\end{aligned}
\label{eqn:MP}
\end{equation}
Here, $\bE^\omega$ and $\bH^\omega$ are for the electric and magnetic fields
for the microwave of angular frequency $\omega$, respectively.
$\bp_{i\mu}$ represents an electric dipole on the $\mu$ sublattice in the $i$th unit cell.
$\bM$ and $\bP$ are operators for the total magnetic moment and total electric dipole, respectively.
For simplicity, we assumed isotropic $g$-factor here.
By the perturbation Hamiltonian $\H'(t)$, magnons at the $\Gamma$ point ($\bk=0$) are excited,
where there are six excited states.
The transition probability to the $\ell$th excited state is given by the Fermi's golden rule as
\begin{align}
W(\ell,\omega)
&= 2\pi \left| \braket{\alpha_{0\ell}^\dagger {\rm GS} | \left( \H'_M + \H'_P \right) | {\rm GS}} \right|^2
        \delta(\omega-E_{0\ell}) \cr
&= 2\pi
\braket{{\rm GS} |  \left( \H'_M + \H'_P \right) \alpha_{0\ell}^\dagger | {\rm GS}} \cr
&~~~~
\times \braket{{\rm GS} | \alpha_{0\ell}  \left( \H'_M + \H'_P \right) | {\rm GS}} \delta(\omega-E_{0\ell}) \cr
&= W_{MM}(\ell,\omega) + W_{EE}(\ell,\omega) + W_{ME}(\ell,\omega).
\end{align}
Here, $|{\rm GS}\rangle$ represents the ground state.
$\alpha_{0\ell}^\dagger$ and $\alpha_{0\ell}$ are creation and annihilation operators
for the $\ell$th excited state at $\bk=0$ whose energy is $E_{0\ell}$.
Notice that $|{\rm GS}\rangle$ is the vacuum state for $\alpha_{\bk\ell}$.
$W_{MM}(\ell,\omega)$ is the transition probability by the magnetic channel,
whereas $W_{EE}(\ell,\omega)$ is by the electric channel.
The former and the latter are caused by the magnetic and electric field components of the microwave, respectively.
$W_{ME}(\ell,\omega)$ represents an interference between the magnetic and electric channels
and leads to nonreciprocal directional dichroism
\cite{Kimura-2020,Miyahara-2011}.
They are given by
\begin{equation}
\begin{aligned}
&W_{MM}(\ell,\omega)
= 2\pi
\braket{{\rm GS} |  \H'_M \alpha_{0\ell}^\dagger | {\rm GS}} \braket{{\rm GS} | \alpha_{0\ell}  \H'_M | {\rm GS}}
\delta(\omega-E_{0\ell}), \cr
&W_{EE}(\ell,\omega)
= 2\pi
\braket{{\rm GS} |  \H'_E \alpha_{0\ell}^\dagger | {\rm GS}} \braket{{\rm GS} | \alpha_{0\ell}  \H'_E | {\rm GS}}
\delta(\omega-E_{0\ell}), \cr
&W_{ME}(\ell,\omega)
= 2\pi
\left( \braket{{\rm GS} |  \H'_M \alpha_{0\ell}^\dagger | {\rm GS}} \braket{{\rm GS} | \alpha_{0\ell}  \H'_E | {\rm GS}} \right. \cr
&~~~~~~~~~~~~~~~~~~~~~
\left.
       + \braket{{\rm GS} |  \H'_E \alpha_{0\ell}^\dagger | {\rm GS}} \braket{{\rm GS} | \alpha_{0\ell}  \H'_M | {\rm GS}} \right) \cr
&~~~~~~~~~~~~~~~~~~~~~~~~
\times \delta(\omega-E_{0\ell}).
\end{aligned}
\label{eqn:Wee}
\end{equation}
Since we focus on the detection of the Higgs amplitude mode,
we do not discuss the nonreciprocal directional dichroism in this paper.

%%%%%%%%%%%%%%%%%%%%%%%%%%%%%%%%%%%%%%%%%%%%%%%%%%%%%%%%%%%%%%%%%%%%%%%%
\subsubsection{Magnetic Channel}
%%%%%%%%%%%%%%%%%%%%%%%%%%%%%%%%%%%%%%%%%%%%%%%%%%%%%%%%%%%%%%%%%%%%%%%%

For the absorption rate of microwave, we calculate the following matrix elements for the magnetic channel:
$\braket{{\rm GS}| \bM \alpha_{0\ell}^\dagger|{\rm GS}}$.
Since we are interested in a one-magnon process, the spin operator in Eq. (\ref{eqn:MP}) can be written as
\begin{align}
\bS_{i\mu\gamma} \rightarrow
\sum_{m=1}^3 \left[ \bS_{0m}(\mu,\gamma) a_{i\mu m} + \bS_{m0}(\mu,\gamma) a_{i\mu m}^\dagger \right].
\label{eqn:one-magnon}
\end{align}
Substituting Eq. (\ref{eqn:one-magnon}) into Eq. (\ref{eqn:MP}), we obtain
\begin{align}
\bM = \sqrt{\frac{N}{2}} g\mu_{\rm B} \sum_\mu \sum_{m=1}^3
\left[ \bS_{0m}(\mu,+) a_{0\mu m} + \bS_{m0}(\mu,+) a_{0\mu m}^\dagger \right].
\label{eqn:Fourier-spin-2}
\end{align}
Here, we used Eq. (\ref{eqn:Fourier-boson}) and represented
$a_{0\mu m}=a_{\bk=0,\mu m}$ and $a_{0\mu m}^\dagger=a_{\bk=0,\mu m}^\dagger$ for $\bk=0$.
$\bS_{mn}(\mu,+) = \bS_{mn}(\mu,l) + \bS_{mn}(\mu,r)$ is the matrix element of the spin operator
of the uniform component of a dimer.

Next, we replace the $a_\bk$ boson with the $\alpha_\bk$ boson by the Bogoliubov transformation.
To perform this, we write the $\ell$th eigenvector of Eq. (\ref{eqn:eigen}) as
\begin{align}
\vec{X}_{\bk \ell} =
\left(
  \begin{array}{c}
    \bm{u}_{\bk \ell} \cr
    \bm{v}_{\bk \ell}
  \end{array}
\right),
\end{align}
where $\bm{u}_{\bk \ell}$ and $\bm{v}_{\bk \ell}$ are 6-dimensional vectors.
These eigenvectors are normalized to satisfy the bosonic commutation relation of $\alpha_{\bk \ell}$.
For the $\ell$th eigenvector, the normalization is given to satisfy
\begin{align}
\sum_{n=1}^{6} \left( \left| u_{\bk \ell n} \right|^2 - \left| v_{\bk \ell n} \right|^2 \right) = 1.
\end{align}
Using $\vec{X}_{\bk \ell}$, we define the following $12\times 12$ matrix:
\begin{align}
\hat{X}_\bk =
\left(
  \begin{array}{cccccc}
    \bm{u}_{\bk 1} & \cdots & \bm{u}_{\bk 6} & \bm{v}_{-\bk 1}^* & \cdots & \bm{v}_{-\bk 6}^* \cr
    \bm{v}_{\bk 1} & \cdots & \bm{u}_{\bk 6} & \bm{u}_{-\bk 1}^* & \cdots & \bm{u}_{-\bk 6}^*
  \end{array}
\right).
\end{align}
The $a_\bk$ boson is then expressed as
\begin{align}
\left(
  \begin{array}{c}
    \bm{a}_\bk \cr
    \bm{a}_{-\bk}^\dagger
  \end{array}
\right)
&= \left( \hat{X}_\bk^{\rm T} \right)^{-1}
\left(
  \begin{array}{c}
    \bm{\alpha}_\bk \cr
    \bm{\alpha}_{-\bk}^\dagger
  \end{array}
\right) \cr
&\equiv
\left(
  \begin{array}{cc}
    \bm{U}_\bk & \bm{V}_\bk \cr
    \bm{V}_{-\bk}^* & \bm{U}_{-\bk}^*
  \end{array}
\right)
\left(
  \begin{array}{c}
    \bm{\alpha}_\bk \cr
    \bm{\alpha}_{-\bk}^\dagger
  \end{array}
\right).
\label{eqn:inverse}
\end{align}
Here, $\bm{U}_\bk$ and $\bm{V}_\bk$ are $6\times 6$ matrices.
$\bm{\alpha}_\bk$ and $\bm{\alpha}_{-\bk}^\dagger$ are 6-dimensional vectors
corresponding to the 6 positive-energy eigenstates of Eq. (\ref{eqn:eigen}).
Substituting Eq. (\ref{eqn:inverse}) into Eq. (\ref{eqn:Fourier-spin-2}), we obtain
\begin{align}
&\braket{{\rm GS}| \bM \alpha_{0\ell}^\dagger |{\rm GS}} \cr
&= \sqrt{\frac{N}{2}} g\mu_{\rm B} \sum_{\mu={\rm A},{\rm B}} \sum_{m=1}^3
\left[ \bS_{0m}(\mu,+) \left( \bm{U}_0 \right)_{m_\mu \ell} \right. \cr
&~~~~~~~~~~~~~~~~~~~~~~~~
\left. + \bS_{m0}(\mu,+) \left( \bm{V}_{0}^* \right)_{m_\mu \ell} \right].
\label{eqn:s-mat}
\end{align}
Here, matrix elements are expressed as $(\cdots)_{m_\mu \ell}$ with $m_\mu$ which depends on the sublattices.
The explicit form of $m_\mu$ is given by $m_{\rm A} = m$ and $m_{\rm B} = m + 3$.

%%%%%%%%%%%%%%%%%%%%%%%%%%%%%%%%%%%%%%%%%%%%%%%%%%%%%%%%%%%%%%%%%%%%%%%%
\subsubsection{Electric Channel}
%%%%%%%%%%%%%%%%%%%%%%%%%%%%%%%%%%%%%%%%%%%%%%%%%%%%%%%%%%%%%%%%%%%%%%%%

In the extended spin-wave theory, we choose the $xyz$ coordinates for the spin space as shown in Fig. \ref{fig:coordinate},
which is the same as the crystal ones.
Then, the direction of the magnetic field is not fixed to the $z$ direction of the spin space.
It can be taken in various directions.
For that reason, the coefficient tensor in Eq. (\ref{eqn:C-pm}) for the electric dipole can be used
under various directions of the field.

As in the magnetic channel, the total electric dipole is expressed as
\begin{align}
\bP = \sqrt{\frac{N}{2}} \sum_\mu \sum_{m=1}^3
\left[ \bp_{0m}(\mu) a_{0\mu m} + \bp_{m0}(\mu) a_{0\mu m}^\dagger \right],
\label{eqn:Fourier-p}
\end{align}
where
\begin{align}
&p_{mn}^\alpha(\mu) = \sum_{\beta=x,y,z} C^\alpha_\beta(\mu) \braket{ i\mu m| \left( \bS_{i\mu l} \times \bS_{i\mu r} \right)_\beta | i\mu n}. \cr
&~~~~~~~~~~~~
(\alpha=x,y,z)
\end{align}
It is independent on the label $i$ for the unit cell.
As in Eq. (\ref{eqn:s-mat}), the matrix element is given by
\begin{align}
&\braket{{\rm GS}| \bP \alpha_{0\ell}^\dagger |{\rm GS}} \cr
&= \sqrt{\frac{N}{2}} \sum_{\mu={\rm A},{\rm B}} \sum_{m=1}^3
\left[ \bp_{0m}(\mu) \left( \bm{U}_0 \right)_{m_\mu \ell} + \bp_{m0}(\mu) \left( \bm{V}_{0}^* \right)_{m_\mu \ell} \right]. \cr
\label{eqn:p-mat}
\end{align}
Substituting Eqs. (\ref{eqn:s-mat}) and (\ref{eqn:p-mat}) into Eq. (\ref{eqn:Wee}),
we can calculate the absorption rate of the microwave (ESR intensity).

%%%%%%%%%%%%%%%%%%%%%%%%%%%%%%%%%%%%%%%%%%%%%%%%%%%%%%%%%%%%%%%%%%%%%%%%

%%%%%%%%%%%%%%%%%%%%%%%%%%%%%%%%%%%%%%%%%%%%%%%%%%%%%%%%%%%%%%%%%%%%%%%%

%%%%%%%%%%%%%%%%%%%%%%%%%%%%%%%%%%%%%%%%%%%%%%%%%%%%%%%%%%%%%%%%%%%%%%%%

\begin{thebibliography}{}
\bibitem{note:intensity}
Let us assume that the spectrum of the resonance peak is described by a distribution function $f(\omega-E(H))$,
which has a peak at $\omega=E(H)$ under a fixed magnetic field $H$.
Here, $\omega$ is the frequency of the microwave, and $E(H)$ represents the energy of the excited state at $H$.
The energy-integrated intensity (ESR intensity) is defined by $I_\omega=\int d\omega f(\omega-E(H))$.
This means that the distribution function $f(x)$ is normalized as $\int dx f(x)=I_\omega$.
On the other hand, the field-integrated intensity is defined by $I_H=\int dH f(\omega-E(H))$.
For a narrow linewidth of the peak, the field dependence of the excitation energy in the vicinity of the resonance
can be expressed with a linear field dependence as $E(H)\simeq \omega + g^* H$ in the integral over $H$.
Here, $g^*$ represents the field gradient of the excitation energy at the resonance.
Then, the field-integrated intensity can be estimated as
$
I_H \simeq \int dH f(\omega-(\omega+g^*H))
= \int dH f(g^*H))
= \frac{1}{g^*} \int d(g^*H) f(g^*H))
= \frac{1}{g^*} \int dx f(x)
= \frac{1}{g^*} I_\omega
$.
From this, we obtain $I_\omega=g^* I_H$.
\end{thebibliography}
\end{document}